\documentclass{svjour3}
\usepackage{enumitem}
\smartqed  
\usepackage{graphicx}
\usepackage{bm}
\usepackage[margin=1in]{geometry}
\usepackage{booktabs}
\usepackage[linesnumbered,ruled,vlined]{algorithm2e}

\SetCommentSty{mycommfont}
\usepackage{algorithmic}
\usepackage{xcolor}
\SetKwInput{KwData}{Data}      
\SetKwInput{KwResult}{Result}  
\SetKwSty{textbf}             
\SetKwFor{For}{for}{do}{end}   
\SetKwComment{Comment}{/* }{ */} 
\DontPrintSemicolon             
\SetKwFunction{Solve}{Solve}     
\SetKw{KwIn}{in}               
\SetKw{KwUsing}{using}         
\SetKw{KwFrom}{from}           
\SetKw{KwWith}{with}         
\RequirePackage{fix-cm}
\usepackage{natbib}
\setcitestyle{numbers,square}
\usepackage{xcolor}
\usepackage{amssymb}
\usepackage{amsmath}
\usepackage{lineno}
\usepackage{setspace}
\usepackage{hyperref}
\usepackage{longtable}
\usepackage{booktabs,subcaption,amsfonts,dcolumn}

\usepackage{algorithmic}
\usepackage{pifont}
\usepackage{siunitx}
\usepackage{booktabs,colortbl,array}
\usepackage{pgfplotstable}
\pgfplotsset{compat=1.8}
\usepackage{multirow}
\usepackage{hhline}
\usepackage{appendix}
\usepackage{enumitem}
\usepackage{booktabs}
\usepackage{tabularx}

\definecolor{rulecolor}{RGB}{0,71,171}
\definecolor{tableheadcolor}{gray}{0.92}

\author{Milad Panahi}

\journalname{Submitted to PNAS Nexus}

\begin{document}
\title{Physics Informed Differentiable Solvers for Learning Parametric Solution Manifolds in Heterogeneous Physical Systems}
\titlerunning{Parameterized PINNs as Differentiable Solvers for Heterogeneous Darcy Flow}
\author{Milad Panahi \and
        Giovanni Michele Porta * \and
         Monica Riva \and
        Alberto Guadagnini}
\date{Received: date / Accepted: date}
\institute{M. Panahi (0000-0002-8776-5297)\and G. M. Porta* (0000-0002-0636-373X) \and A. Guadagnini (0000-0003-3959-9690) \and M. Riva (0000-0002-7304-4114) \at Dipartimento di Ingegneria Civile e Ambientale, Politecnico di Milano, Piazza L. da Vinci 32, 20133 Milano, Italy \\
    * To whom correspondence should be addressed: \email{giovanni.porta@polimi.it}
}

\maketitle

\begin{abstract}
Learning the full family of solutions to parameterized partial differential equations (PDEs) is a central challenge to our ability to model the behavior of heterogeneous systems, with a variety of fundamental and application-oriented implications in fields such as hydrogeology where system properties exhibit significant (and often uncertain) spatial heterogeneity. We address this by reformulating a Physics-Informed Neural Network (PINN) as a differentiable solver that learns the continuous solution manifold for steady-state Darcy flow. Our framework requires only a single training run, circumventing the need for costly re-training for each new parameter instance. Its versatility is demonstrated through two representations of spatially heterogeneous hydraulic conductivity fields: a direct analytical form and a novel data-driven formulation resting on an autoencoder to create a low-dimensional latent encoding. A key innovation is the integration of the differentiable decoder into the physics-informed loss function, enabling on-the-fly reconstruction of complex conductivity fields via automatic differentiation. The approach yields accurate, mass-conserving flow solutions and supports efficient uncertainty quantification, providing a general methodology for physics-constrained data-driven modeling of heterogeneous systems.

\section*{Article Highlights}
\begin{itemize}
    \item[\ding{70}] A parameterized PINN framework is presented as a differentiable solver that learns the entire solution manifold for heterogeneous Darcy flow.
    \item[\ding{70}] A novel autoencoder-based parameterization is introduced, where the differentiable decoder is integrated into the physics-informed loss to reconstruct complex conductivity fields on-the-fly.
    \item[\ding{70}] The trained surrogate is shown to be accurate, computationally efficient, and to robustly preserve local mass conservation without being explicitly trained on this requirement.
    \item[\ding{70}] The rapid inference speed of the learned solver makes ensemble-based analyses, such as Monte Carlo uncertainty quantification, computationally tractable.
\end{itemize}


\section*{Significance Statement}
Modeling the behavior of complex physical systems, from groundwater flow to material science, is often hindered by the high computational cost of solving the governing parameterized equations for every new set of conditions. We overcome this bottleneck by framing a physics-informed neural network as a $differentiable solver$ capable of learning the entire family of solutions in a single training run. Our framework uniquely integrates a differentiable generative model directly into a physics-informed loss, enabling it to solve systems with complex, spatially heterogeneous properties. We demonstrate that this purely physics-trained model robustly preserves physical principles such as local mass conservation. This work provides a flexible and efficient pathway for building physically consistent $digital twins$ for uncertainty quantification in a broad range of applications.
\end{abstract}

\section{INTRODUCTION}
\label{sec:introduction}
Describing our conceptual understanding of the behavior of a physical system through a set of partial differential equations (PDEs) is at a key element of modern science and engineering. In several critical domains (ranging from climate science to hydrogeology or materials science) formulations of governing equations typically embed a number of system parameters \citep{rackauckas2020universal, lavin2021simulation}. With reference to the hydrogeology context, for example, accurate simulation of subsurface fluid flow is challenged by the spatial heterogeneity exhibited by hydraulic conductivity, $K(\mathbf{x})$. The latter can span several orders of magnitude across a given geologic system and is characterized by complex (often multi-scale) patterns, markedly imprinting the overall system dynamics, including the emergence of preferential flow pathways \citep{sanchez1996scale, neuman1990universal, bear2013dynamics, rubin2003applied}. Accurate characterization of the uncertainty associated with these parameters and its rigorous propagation through predictive models is at the heart of reliable risk assessment and uncertainty-aware decision-making \citep{tartakovsky2013assessment, janetti2021natural, riva2015probabilistic}. Traditionally, this has been achieved upon relying on collections (or ensembles) of forward model simulations (such as Monte Carlo methods \citep{ballio2004convergence}) implemented through the use of numerical solvers (such as, e.g., Finite Element Methods (FEM)). Yet, as the dimensionality of the parameter space increases, these approaches become computationally prohibitive. This limitation poses a major challenge for modeling complex natural systems, where high-dimensional parameterizations are important to capture heterogeneity and process interactions, ultimately constraining our ability to quantify and manage risk under uncertainty.

Beyond parameter uncertainty, additional challenges stem from model structural errors and the inherent stochasticity of natural systems. Capturing the combined effects of these uncertainty sources in a unified framework remains computationally demanding, thus underscoring the need for efficient and scalable strategies to quantify uncertainty in complex environmental models.

Recent advances in machine learning applications within hydrological sciences have reignited a long-standing debate over the roles of traditional process-based models and purely data-driven approaches \citep{beven1987towards, lecun2015deep, mohri2018foundations}. Process-based models, grounded in established physical laws, provide interpretability and physical consistency but often rely on simplified assumptions that can limit predictive accuracy. Modern deep learning models can, in some contexts, achieve a somehow superior performance by capturing complex, nonlinear relationships directly from observational data, particularly in settings where sparse (and scarce) data hinder robust calibration of physical models \citep{zhu2019physics, asher2015review, gupta2006model}. This evolving landscape in the age of artificial intelligence underscores the need for hybrid modeling strategies that bridge these two paradigms upon integrating physical understanding with data-driven flexibility \citep{nearing2021role, beven2020deep, mishra2018machine}. The documented potential of such a data-driven perspective is giving rise to a prominent direction in Scientific Machine Learning (SciML) \citep{willard2022integrating, Blechschmidt}, i.e., \textit{neural operator learning} \citep{kovachki2023neural}, featuring models such as Fourier Neural Operators (FNOs) \citep{li2020fourier} and DeepONets \citep{lu2021learning} aimed at learning the way one can map a space of input parameters onto the corresponding space of a target solution representing the system behavior. An alternative data-driven strategy explicitly tackles the high-dimensionality of parameter fields through dimensionality reduction. For instance, the Karhunen-Lo\`{e}ve deep learning (KL-DNN) method \citep{wang2025karhunen} uses Karhunen-Lo\`{e}ve expansions to project parameter fields and system state variables into low-dimensional latent spaces, subsequently learning an efficient mapping between these reduced coordinates \citep{CHEN2021110666, liu20213d, he2011enhanced, tang2021deep}. This approach of learning on a reduced-order manifold is conceptually similar to the autoencoder-based parameterization we explore in this work. It is otherwise important to note that all of these powerful frameworks (both direct operator learners and latent-space models such as KL-DNN) remain fundamentally data-driven. They are trained in a supervised manner on large, pre-computed datasets of input-output pairs. While they have shown remarkable success in their intended applications, for example in accelerating simulations for computational fluid dynamics \citep{li2020fourier} and data-driven subsurface flow modeling \citep{tang2021deep}, they often face challenges with out-of-distribution generalization and demand a significant upfront investment in high-fidelity simulations \citep{lu2022comprehensive}. In contrast, Physics-Informed Neural Networks (PINNs) offer an approach that circumvents this reliance on pre-computed data by embedding the physical laws directly into the network loss function \citep{lagaris1998artificial, raissi2019physics, karniadakis2021physics, raissi2024physics, luo2025physics}.

This work addresses a critical methodological trade-off in Scientific Machine Learning: the balance between the \textit{solution-specific accuracy} of PINNs and the \textit{operator-level generalization} of Physics-Informed Neural Operators (PINOs) \citep{li2024physics, noguer2025mathematics}. While PINOs offer efficiency in parametric studies, they often function as regularized data-driven models, potentially sacrificing the pinpoint, resolution-independent fidelity of direct solvers. Conversely, extending the rigorous PINN framework to learn a continuous solution manifold (effectively turning it into a parameterized solver) is non-trivial. Such deep, parameterized networks are prone to severe training pathologies, notably ``spectral bias''—the tendency to over-smooth high-frequency features \citep{Wang_2022, rahaman2019spectral, wang2021eigenvector}—and stiff optimization landscapes plagued by unbalanced gradients \citep{wang2022and, wang2021understanding, mcclenny2023self}. 

To overcome these barriers, we ground our approach in the principles of differentiable programming \citep{baydin2018automatic}, tasking the network to approximate the solution manifold defined implicitly by the governing equations \citep{amos2017optnet, fletcher2000practical}. We demonstrate that achieving this requires moving beyond standard architectures by employing adaptive residual connections (PirateNets) \citep{wang2024piratenets, he2016deep, anagnostopoulos2023residual} and advanced training strategies like curriculum learning \citep{krishnapriyan2021characterizing, penwarden2023metalearning} or meta-learning \citep{qin2022metapde, CHEN2022110996, Liu_2022, pmlrv70finn17a, antoniou2019trainmaml, nichol2018reptile}.

At their core, PINNs leverage concepts underlying Implicit Neural Representations (INRs), where a neural network learns to represent a signal as a continuous function of its coordinates \citep{kingma20153rd, sitzmann2020implicit, tancik2020fourier, mildenhall2021nerf}. While initially prominent in computer vision, the application of INRs to represent dynamic physical systems is a growing area of research \citep{yin2022continuous, raissi2023open}. A PINN can thus be viewed as an INR for the solution field (i.e., the set of physical quantities, such as pressure or velocity, whose behavior is described by the PDE) \citep{dashtbayaz2025physics}. Network parameters are optimized to satisfy the governing physical laws by embedding the PDE residuals directly into the loss function, rather than fitting pre-existing data. In principle, such a physics-informed approach is applicable even in the complete absence of measurement data, thus significantly reducing the need for relying on large datasets of pre-computed input-output simulation pairs (often referred to as \textit{labeled data}). However, the standard PINN framework, as originally formulated \citep{raissi2019physics}, is designed to solve a single PDE instance and is not inherently built for parameterized systems. Consequently, a straightforward application to a parameterized problem would require a complete (and costly) re-training for each new set of parameter values (e.g., when considering diverse values for material properties or boundary conditions). This re-introduces the very computational bottleneck the approach was designed to circumvent.

In this broad framework, two distinct paradigms have been developed in scientific machine learning: (a) the solution-specific approach of PINNs and (b) the data-driven, operator-learning approach of Neural Operators (e.g., FNOs, DeepONets). These are not mutually exclusive and have inspired powerful hybrid methods. Notably, \textit{Physics-Informed DeepONets} (PI-DeepONets) \citep{goswami2023physics, wang2021learning} and \textit{Physics-Informed Neural Operators} (PINOs) \citep{li2024physics} augment their respective data-driven frameworks with physics-based loss functions. This augmentation alleviates data dependency and improves physical consistency of the learned operator. Otherwise, our work adheres to the core PINN philosophy by treating the entire parameterized system as an end-to-end \textit{differentiable program} \citep{NEURIPS2018_DP, innes2019differentiable}. This view aligns with the paradigm of \textit{differentiable physics}, where the goal is to learn from the gradients associated with a physics-based simulator or solver itself rather than to merely regularize a data-driven model through a suitably designed loss function \citep{um2020solver, thuerey2021physics, NEURIPS2018_DP, Karnakov2024diffinv}. In this view, the neural network is actually the solver rendering system states that satisfy desired physics-based constraints. Upon parameterizing the neural network through a parameter vector $\boldsymbol{\lambda}$ and ensuring differentiability throughout the entire computational graph (i.e., from parameters through physics residuals), one can directly learn an approximation of the solution manifold. This design establishes our framework as a fully physics-constrained approach with operator-like learning capabilities. Hence, our study is intended to complement existing methods, thereby enhancing their potential for fundamental research and practical applications.

Our study is set in the context of the growing body of research focused on physics-based unsupervised learning and surrogate probabilistic modeling. It aims at enabling integration of constraints stemming from our conceptual understanding of the main physics underlying the system behavior directly into learning architectures for the simulation of a given process under uncertainty. In this sense, a distinctive objective of our work is to advance the paradigm of parameterized PINNs upon framing these as \textit{differentiable solvers} that learn the entire \textit{solution bundle} \citep{flamant2020solving} (i.e., the manifold of solutions in high-dimensional spaces) across the entire range of (uncertain) parameter values (as encapsulated in $\boldsymbol{\lambda}$), as opposed to providing a single solution for a fixed instance of $\boldsymbol{\lambda}$.

This approach represents a significant departure from traditional uncertainty quantification workflows. Standard Monte Carlo methods typically generate a large ensemble of solution samples, from which statistical moments (e.g., mean and (co)variance) are computed to characterize the output distribution under uncertain inputs. By learning the entire solution manifold, our framework provides a complete representation of uncertainty. The resulting trained model can be queried rapidly for any parameter instance. Hence, it allows for the reconstruction of the full solution distribution, not being limited to its low-order moments.

We build on our previous work with PINNs for Uncertainty Quantification (PINN-UU) \citep{panahi2025modeling}, which extends the approaches of \citep{ARTHURS2021110364} and \cite{HyperPINN2021Filipe, ha2016hypernetworks} for solving parametrized PDEs through explicit and implicit parameter encoding strategies, respectively. Here, we advance the PINN-UU framework by introducing a unified neural solver that takes both physical coordinates $\mathbf{x}$ and system parameters $\boldsymbol{\lambda}$ as direct inputs, similar to \citep{ARTHURS2021110364}. This design enables one to integrate explicit parameter inputs with implicit encoding of (spatially) heterogeneous PDE parameters as latent vectors through manifold learning in the Meta-Auto-Decoder framework \citep{huang2022meta}.

A central contribution of our work is its extension to systems where the parameter vector $\boldsymbol{\lambda}$ encodes, in a low-dimensional latent form, a complex heterogeneous field \citep{delhomme1979spatial, dell2019solute, ye2004nonlocal}, that is often learned via generative models such as Variational Autoencoders \citep{tait2020variational}. Within an operator-learning perspective, this formulation requires differentiating through the decoding map that reconstructs the full spatial field rather than not only modeling the system response to the latent parameters but also differentiating through the decoding map that reconstructs the full spatial field. Incorporating this differentiable decoding into the physics-informed loss positions our approach within the broader paradigm of textit{differentiable physics} \citep{thuerey2021physics, um2020solver}, enabling training that is consistent with latent structure as well as physical constraints.

We apply our theoretical and operational framework to a steady-state Darcy flow scenario, where the parameter vector $\boldsymbol{\lambda}$ encodes a spatially heterogeneous hydraulic conductivity field. 
The key novelties in our study are listed in the following.
\begin{enumerate}[label=(\roman*)]
    \item Our work tackles simulation in spatially heterogeneous materials through PINNs. To this end we adopt a network architecture inspired by PirateNets \citep{wang2024piratenets}, leveraging adaptive residual connections to promote stable and efficient training. We then demonstrate the effectiveness of this architecture as a parameterized solver for the hydraulic head and velocity fields associated with a mathematical description of groundwater flow based on Darcy’s law.
    \item We provide simulation tool capable of handling input parameters characterized by given distribution. Practical applications motivating our study are forward quantification of uncertainty, sensitivity analysis, experimental design. We test wo distinct methods for constructing the differentiable mapping from the parameters to model outputs: (1) a direct, analytical (functional) parameterization, and (2) a flexible, data-driven approach using a pre-trained, coordinate-based autoencoder that approximates inputs characterized by spatial complexity.
    \item We consider simulation in a high-dimensional input domain featuring both spatial coordinates and the parameter space. To this end we develop a multi-stage transfer learning strategy to accelerate convergence and improve robustness of the solutions across the high-dimensional domain of the spatially distributed conductivities. 
\end{enumerate}

The remainder of the paper is organized as follows. Section~\ref{sec:methodology} details the mathematical formulation of the parameterized Darcy flow scenario and our proposed methodology. Section~\ref{sec:results} illustrates and discusses the ensuing numerical results. Finally, Section~\ref{sec:conclusion} provides key conclusions and outlines future research directions.

\section{Methodology: Differentiable Physics-Informed Learning Framework}
\label{sec:methodology}
Here, we illustrate our differentiable physics-informed learning framework. The neural network architecture and embedded physical constraints are integrated into a single computational graph, enabling simultaneous adherence to data and physics-constrained optimization via gradient-based methods. We detail the formal problem statement, the physics-informed learning framework, our strategies for encoding spatially heterogeneous conductivities, and the training approach.

\subsection{\textbf{A General Framework for Parameterized PDEs}}
\label{sec:general_framework}
Let the spatial domain be an open set $\Omega \subset \mathbb{R}^d$. The PDE system governing the system behavior is characterized by a vector of parameters $\boldsymbol{\lambda}$ of dimension ${d_\lambda}$ defined within a parameter space $\Lambda \subset \mathbb{R}^{d_\lambda}$. 

The relationship between the parameters and the solution is implicitly defined by a system of differential operators. We define an operator $\mathcal{F}[u(\mathbf{x},\boldsymbol{\lambda})] = \mathcal{F}_{\mathbf{x},\boldsymbol{\lambda}}(u)$ that represents the governing equations within the spatial domain $\Omega$ and the parameter space $\Lambda$, and an operator $\mathcal{B}[u(\mathbf{x},\boldsymbol{\lambda})]=\mathcal{B}_{\mathbf{x},\boldsymbol{\lambda}}(u)$ for the boundary conditions on $\partial\Omega$. The vector $\mathbf{x}$ represents the spatial coordinates. The framework presented in this study is focused on steady-state problems, where the solution $\mathbf{u}$ does not evolve in time.

The parameterized PDE problem is to find a solution $\mathbf{u} \in \mathcal{S}$ for each given parameter instance $\boldsymbol{\lambda} \in \Lambda$ such that:
\begin{align}
    \mathcal{F}_{\mathbf{x},\boldsymbol{\lambda}}(u) &= f_{\mathbf{x},\boldsymbol{\lambda}}, \quad \forall \mathbf{x} \in \Omega \\
    \mathcal{B}_{\mathbf{x},\boldsymbol{\lambda}}(u) &= b_{\mathbf{x},\boldsymbol{\lambda}}, \quad \forall \mathbf{x} \in \partial\Omega
\end{align}
where the parameter vector $\boldsymbol{\lambda}$ drives the solution by affecting the operators themselves as well as the source terms here defined as $f_{\mathbf{x},\boldsymbol{\lambda}}$ and $b_{\mathbf{x},\boldsymbol{\lambda}}$. While the formulation above defines the problem for individual parameter values, our primary goal is to learn the \textit{solution map}, $\mathcal{G}: \Lambda \to \mathcal{S}$. This map takes any given parameter vector $\boldsymbol{\lambda}$ from the parameter space and returns the corresponding solution $\mathbf{u} = \mathcal{G}(\boldsymbol{\lambda})$. We assume here that for any given $\boldsymbol{\lambda}$ in the considered parameter space, the problem is well-posed and admits a unique solution.

We approximate such solution map upon relying on a single, unified neural network, $\mathcal{N}$, parameterized by trainable weights and biases $\boldsymbol{\theta}$. This network acts as a differentiable solver by learning a function that maps jointly spatial coordinates and parameter values to the solution field:
\begin{equation}
    \hat{u}(\mathbf{x}, \boldsymbol{\lambda};\boldsymbol{\theta} ) = \mathcal{N}(\mathbf{x}, \boldsymbol{\lambda}; \boldsymbol{\theta}).
    \label{eq:nn_ansatz}
\end{equation}
Effectively, the network $\mathcal{N}$ represents an approximation $\mathcal{G}_{\boldsymbol{\theta}} \approx \mathcal{G}$ of the solution map. The training process aims at discovering the optimal network parameter vector $\boldsymbol{\theta}^*$ minimizing the residuals of the governing equations, which are scalar functions defined as:
\begin{align}
    r_{pde}(\mathbf{x}, \boldsymbol{\lambda}; \boldsymbol{\theta}) &:= \mathcal{F}[\hat{u}(\mathbf{x},\boldsymbol{\lambda};\boldsymbol{\theta})] - f(\mathbf{x},\boldsymbol{\lambda}), \\
    r_{bc}(\mathbf{x}, \boldsymbol{\lambda} ; \boldsymbol{\theta})&:= \mathcal{B}[\hat{u}(\mathbf{x},\boldsymbol{\lambda};\boldsymbol{\theta})] -b(\mathbf{x},\boldsymbol{\lambda}).
\end{align}
The composite loss function $\mathcal{J}(\boldsymbol{\theta})$, which is employed to constrain $\boldsymbol{\theta}$, aggregates the mean squared residuals associated with sets of collocation points. Let $\mathcal{T}_{pde} = \{(\mathbf{x}_i, \boldsymbol{\lambda}_i)\}_{i=1}^{N_{pde}}$ be a set of $N_{pde}$ collocation points sampled from the domain space $\Omega \times \Lambda$, and $\mathcal{T}_{bc} = \{(\mathbf{x}_j, \boldsymbol{\lambda}_j)\}_{j=1}^{N_{bc}}$ be a set of $N_{bc}$ collocation points from the boundary space $\partial\Omega \times \Lambda$. The loss function is then:
\begin{equation}
    \mathcal{J}(\boldsymbol{\theta}) = \frac{1}{N_{pde}} \sum_{i=1}^{N_{pde}} \left( r_{pde}(\mathbf{x}_i, \boldsymbol{\lambda}_i; \boldsymbol{\theta}) \right)^2
                      + \frac{1}{N_{bc}} \sum_{j=1}^{N_{bc}} \left( r_{bc}(\mathbf{x}_j, \boldsymbol{\lambda}_j; \boldsymbol{\theta}) \right)^2,
    \label{eq:general_loss}
\end{equation}
Starting from these definitions, we introduce in the following the specific problem formulation adopted to solve the considered physical problem, corresponding to fluid flow across a saturated heterogeneous porous medium.

\subsection{\textbf{Problem Formulation: The Parameterized Darcy System}}
\label{sec:problem_formulation}
We consider the problem of steady-state, single-phase fluid flow in a two-dimensional porous medium. The setup is designed to be representative of a laboratory-scale experiment and the problem is formulated and solved on a dimensionless unit square domain, $\Omega = [0, 1] \times [0, 1]$. All quantities presented are therefore dimensionless, though they can be scaled to a consistent set of physical units for a specific application.

The system is governed by the continuity equation (i.e., mass conservation) and Darcy's law \citep{neuman1977theoretical}. The domain boundary $\partial\Omega$ is partitioned into Dirichlet ($\Gamma_D$) and Neumann ($\Gamma_N$) portions. Our primary goal is to find the solution pair $\mathbf{u}(\mathbf{x}) = (h(\mathbf{x}), \mathbf{v}(\mathbf{x}))$, consisting of dimensionless hydraulic head and the fluid velocity vector. The solution depends on a parameter vector $\boldsymbol{\lambda} \in \Lambda$ that defines the spatially heterogeneous hydraulic conductivity, $K(\mathbf{x}; \boldsymbol{\lambda})$. We consider here an isotropic porous medium. Hence hydraulic conductivity can be described as a scalar quantity.

The governing equations, often referred to as the strong form of the PDE system, are given for a specific $\boldsymbol{\lambda}$ as:
\begin{subequations}
\label{eq:full_darcy_system}
\begin{align}
    & \nabla \cdot \mathbf{v}(\mathbf{x}; \boldsymbol{\lambda}) + f(\mathbf{x}) = 0, && \forall \mathbf{x} \in \Omega \label{eq:strong_cont_method} \\
    & \mathbf{v}(\mathbf{x}; \boldsymbol{\lambda}) + K(\mathbf{x}; \boldsymbol{\lambda}) \nabla h(\mathbf{x}; \boldsymbol{\lambda}) = \mathbf{0}, && \forall \mathbf{x} \in \Omega \label{eq:strong_darcy_method} \\
    & h(\mathbf{x}; \boldsymbol{\lambda}) = g_D(\mathbf{x}), && \forall \mathbf{x} \in \Gamma_D \label{eq:dirichlet_bc} \\
    & \mathbf{v}(\mathbf{x}; \boldsymbol{\lambda}) \cdot \mathbf{n} = g_N(\mathbf{x}), && \forall \mathbf{x} \in \Gamma_N \label{eq:neumann_bc}
\end{align}
\end{subequations}
where $f(\mathbf{x})$ is a source term (considered to be zero in this study), and $\mathbf{n}$ is a (outward-pointing) unit vector normal to the domain boundary. Functions $g_D$ and $g_N$ represent prescribed boundary values. Our particular set of boundary and initial conditions is inspired by the experimental work in \citep{MAINA201855}. For our specific setup:

\begin{itemize}
    \item[\ding{68}] The Dirichlet boundary, $\Gamma_D = \{\mathbf{x} \in \partial\Omega \,|\, x=1, 0.8 < y < 1 \}$, represents an outlet with a zero-head condition, such that $g_D=0$\,m.
    \item[\ding{68}] The Neumann boundary ($\Gamma_N = \partial\Omega \setminus \Gamma_D$) is further composed of:

    \begin{itemize}
        \item[\ding{67}] An inlet section at $\Gamma_{N, \text{in}} = \{\mathbf{x} \in \partial\Omega \,|\, y=0, 0 < x < 0.2 \}$, where the influx is prescribed by a parabolic profile with a peak velocity of $q_{\text{max}} = 1.0\,\text{m/d}$:
        \begin{equation}
            g_N(x,y) = q_{\text{max}} \cdot 4 \left( \frac{x}{0.2} \right) \left( 1 - \frac{x}{0.2} \right).
            \label{eq:inlet_profile}
        \end{equation}
        \item[\ding{67}] Impermeable walls, $\Gamma_{N, \text{wall}} = \Gamma_N \setminus \Gamma_{N, \text{in}}$, where the normal flux is zero ($g_N=0$\,m/d).
    \end{itemize}
\end{itemize}
The main objective of the work is to learn an approximation of the solution operator ($\mathcal{G}: \Lambda \to \mathcal{V}$) that maps any instance of parameter vector $\boldsymbol{\lambda}$ to its associated unique solution $\mathbf{u}$ upon satisfying system \eqref{eq:full_darcy_system}.

\subsection{\textbf{The Physics-Informed Learning Framework}}
\label{sec:pinn_framework}
\begin{figure}[!ht]
    \centering
    \includegraphics[width=0.9\textwidth]{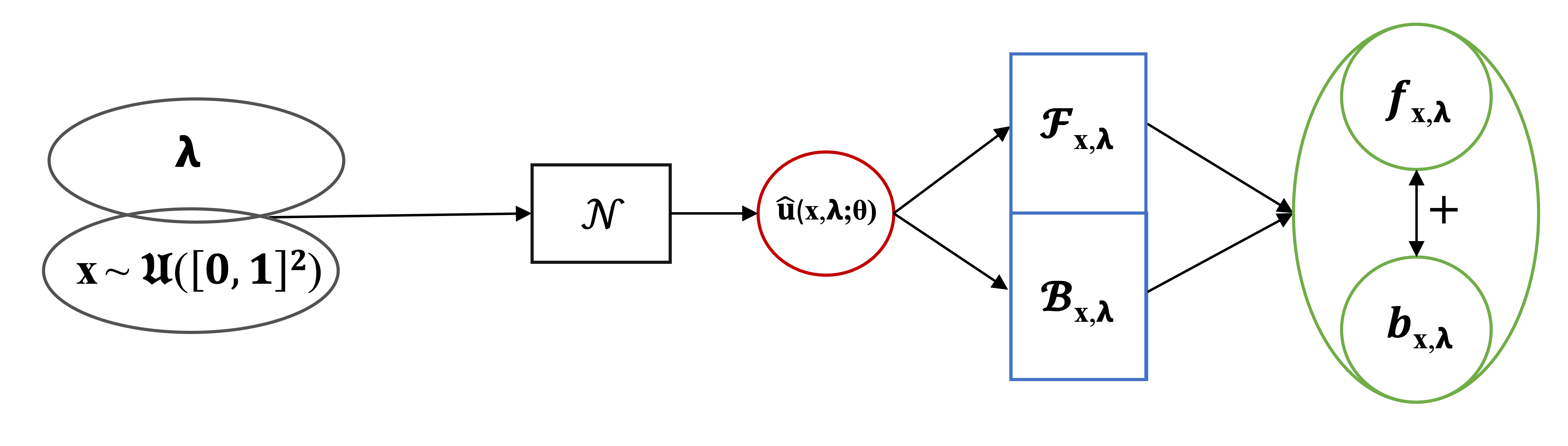}
    \caption{Schematic depiction of the parameterized PINN framework serving as a differentiable solver. The network $\mathcal{N}$ takes both spatial coordinates $\mathbf{x}$ and a generic instance of parameter vector $\boldsymbol{\lambda}$ as input to approximate the solution field $\hat{u}(\mathbf{x}, \boldsymbol{\lambda};\mathbf{\theta})$. This architecture forms the basis for learning the entire solution manifold.}
    \label{fig:pinn_ansatz_schematic}
\end{figure}

Design and formulation of a Physics-Informed Neural Network (PINN) functioning as a \textit{differentiable solver} are at the core of our methodology. As illustrated in Figure~\ref{fig:pinn_ansatz_schematic}, we use a single neural network that jointly processes both spatial and parameter inputs. It learns mapping from the combined spatial domain and parameter space directly onto the solution manifold. The network takes space coordinates $\mathbf{x}$ and a parameter vector $\boldsymbol{\lambda}$ as inputs and renders the corresponding solution field ($\mathbf{u}$) as its output. This architecture enables a single trained model to represent the entire family of solutions across the parameter space $\Lambda$.

\subsubsection{\textbf{The Neural Solver Ansatz}}
We approximate the solution operator by learning the scalar hydraulic head field directly. A single, unified neural network, $\mathcal{N}$, parameterized by trainable weights and biases (forming the entries of vectore $\boldsymbol{\theta}$), is defined to map spatial coordinates and parameters to head values corresponding to network outputs:
\begin{equation}
    \hat{h}(\mathbf{x},\boldsymbol{\lambda}; \boldsymbol{\theta}) = \mathcal{N}(\mathbf{x}, \boldsymbol{\lambda}; \boldsymbol{\theta}).
    \label{eq:nn_ansatz1}
\end{equation}
The ensuing velocity field, $\hat{\mathbf{v}}$, is not a direct output of the network. Instead, it is obtained from the head output by applying automatic differentiation (AD) \citep{baydin2018automatic} to enforce momentum balance, as embedded in Darcy's law (Eq. \ref{eq:strong_darcy_method}):
\begin{equation}
    \hat{\mathbf{v}}(\mathbf{x}; \boldsymbol{\lambda}) = -K(\mathbf{x}; \boldsymbol{\lambda}) \nabla \hat{h}(\mathbf{x}; \boldsymbol{\lambda}).
    \label{eq:velocity_derivation}
\end{equation}
This formulation ensures that the learned velocity field is, by construction, divergence-free with respect to the learned head field, the latter being a key aspect of the considered physical problem. We employ a network architecture inspired by PirateNets \citep{wang2024piratenets}, which is here selected for its stability (details are in Supporting Information ~\ref{app:piratenet}).

\subsubsection{The Physics-Informed Loss Function}
Values of network parameters $\boldsymbol{\theta}$ are optimized by minimizing a composite loss function (denoted as $\mathcal{J}(\boldsymbol{\theta})$) constructed from the residuals of the governing physical laws. This loss function comprises distinct terms for the PDE residual within the domain and for each type of boundary condition. The core principle is that these physical laws must be satisfied across the whole spatial domain $\mathbf{x} \in \Omega$ and for the whole parameter space $\boldsymbol{\lambda} \in \Lambda$. Hence, we define the joint spatio-parameter domain as $\Omega_{\boldsymbol{\lambda}} = \Omega \times \Lambda$.

\paragraph{\textbf{Domain Residuals (with respect to Mass conservation and Darcy’s Law).}} Residuals enforce the governing physical laws throughout the domain $\Omega_{\boldsymbol{\lambda}}$, which includes both the spatial and parametric domains. These are defined as:
\begin{equation}
    r_{\text{cont}}(\mathbf{x}, \boldsymbol{\lambda}; \boldsymbol{\theta}) := \nabla \cdot \hat{\mathbf{v}}(\mathbf{x}; \boldsymbol{\lambda}) + f(\mathbf{x}), \quad \forall (\mathbf{x}, \boldsymbol{\lambda}) \in \Omega_{\boldsymbol{\lambda}} \\
\end{equation}

\paragraph{\textbf{Boundary Residuals.}} Boundary residuals enforce the specified conditions on the domain boundary, $\partial\Omega_{\boldsymbol{\lambda}}$. We define distinct residual functions for Dirichlet and Neumann conditions. These may then be applied to diverse parts of the boundary, $\partial\Omega_{\boldsymbol{\lambda}, D}$ and $\partial\Omega_{\boldsymbol{\lambda}, N}$, respectively.
\begin{itemize}
    \item[\ding{68}] Dirichlet Residual ($r_D$): For a prescribed head condition $h(\mathbf{x}) = g_D(\mathbf{x})$ on $\partial\Omega_D$, the residual measures the difference between the network output and the target value:
    \begin{equation}
        r_{D}(\mathbf{x}, \boldsymbol{\lambda}; \boldsymbol{\theta}) := \hat{h}(\mathbf{x}; \boldsymbol{\lambda}) - g_D(\mathbf{x}), \quad \forall (\mathbf{x}, \boldsymbol{\lambda}) \in \partial\Omega_{\boldsymbol{\lambda}, D}
    \end{equation}
    \item[\ding{68}] Neumann Residual ($r_N$): For a prescribed flux condition on $\partial\Omega_N$, the residual measures the difference between the component of flux normal to the boundary and its target counterpart, $g_N(\mathbf{x})$. The flux is computed via AD:
    \begin{equation}
        r_{N}(\mathbf{x}, \boldsymbol{\lambda}; \boldsymbol{\theta}) := \hat{\mathbf{v}}(\mathbf{x}; \boldsymbol{\lambda}) \cdot \mathbf{n}(\mathbf{x}) - g_N(\mathbf{x}), \quad \forall (\mathbf{x}, \boldsymbol{\lambda}) \in \Omega_{\boldsymbol{\lambda}, N}
    \end{equation}
    where $\mathbf{n}(\mathbf{x})$ is the (outward) unit vector normal to the boundary.
\end{itemize}

\paragraph{Composite Loss Function.} Following \eqref{eq:general_loss}, the total loss function $\mathcal{J}(\boldsymbol{\theta})$ is the weighted sum of the mean squared residuals described above. In practice, residuals (or errors) are approximated by sampling $N_r$ collocation points from the domain $\Omega \times \Lambda$, and $N_{bD}$ and $N_{bN}$ points from the Dirichlet and Neumann boundary segments, respectively. The individual loss components are:
\begin{align}
    \mathcal{J}_{domain} = &\frac{1}{N_r} \sum_{i=1}^{N_r} (r_{\text{cont}}(\mathbf{x}_i, \boldsymbol{\lambda}_i; \boldsymbol{\theta}))^2, \\
    \mathcal{J}_{boundary} = \frac{1}{N_{bD}} \sum_{j=1}^{N_{bD}} &(r_{D}(\mathbf{x}_j, \boldsymbol{\lambda}_j; \boldsymbol{\theta}))^2 + \frac{1}{N_{bN}} \sum_{k=1}^{N_{bN}} (r_{N}(\mathbf{x}_k, \boldsymbol{\lambda}_k; \boldsymbol{\theta}))^2. \\
\end{align}
These components are then combined to form the total loss:
\begin{equation}
    \mathcal{J}(\boldsymbol{\theta}) = w_{d}\mathcal{J}_{domain} + w_{b}\mathcal{J}_{boundary}.
    \label{eq:total_loss_detailed}
\end{equation}
The loss weights, $w_d$ and $w_b$, are handled dynamically during optimization to ensure stable training, as described in the fllowing.

\subsubsection{Gradient Balancing for Stable Training}
\label{BoundedGradNorm}

A known pathology in training PINNs is the difficulty in balancing contributions of different loss components \cite{wang2021understanding}. A large gradient from one term can overwhelm the others, leading to poor convergence. To address this issue, we employ "Bounded GradNorm", an adaptive gradient balancing scheme inspired by GradNorm \cite{chen2018gradnorm}.

At each training step $t$, we compute the gradients of the individual loss components with respect to the network parameters $\boldsymbol{\theta}$:
\begin{equation}
    \mathbf{g}_{\text{domain}}^t = \nabla_{\boldsymbol{\theta}} \mathcal{J}_{\text{domain}} \quad \text{and} \quad \mathbf{g}_{\text{boundary}}^t = \nabla_{\boldsymbol{\theta}} \mathcal{J}_{\text{boundary}}.
\end{equation}
We then compute the L2 norm of these gradient vectors. The ratio of these norms at the current training step, $r_t = \|\mathbf{g}_{\text{boundary}}^t\|_2 / (\|\mathbf{g}_{\text{domain}}^t\|_2 + \epsilon)$, where $\epsilon$ is a small constant required for numerical stability, is used to update an exponential moving average (EMA) of the ratio, denoted as $\zeta_t$:
\begin{equation}
    \zeta_t = (1-\beta) \zeta_{t-1} + \beta r_t,
\end{equation}
Here, $\zeta_0$ is initialized to 1.0 and $\beta \in (0, 1)$ is a weighting factor ($\beta=0.1$ in our implementation). To ensure stability, this moving average is clipped to a predefined range $[\zeta_{\min}, \zeta_{\max}]$, yielding a bounded ratio $\zeta_t^{\text{clipped}}$.

From this bounded ratio, we define the final scaling factors for the domain and boundary gradients (denoted as $\beta_d$ and $\beta_b$, respectively) as follows:
\begin{equation}
    \beta_d = \max(\zeta_t^{\text{clipped}}, 1.0) \quad \text{and} \quad \beta_b = \max(1.0 / \zeta_t^{\text{clipped}}, 1.0).
\end{equation}
This formulation ensures that one scaling factor is always 1.0 while the magnitude of the other is adjusted on the basis of the largest gradient, thus effectively maintaining a balance between their influences. The final parameter update is then performed using these scaled gradients:
\begin{equation}
    \Delta\boldsymbol{\theta}^t = -\eta \left( \beta_d \mathbf{g}_{\text{domain}}^t + \beta_b \mathbf{g}_{\text{boundary}}
    ^t \right),
\end{equation}
where $\eta$ is the learning rate.

By dynamically re-scaling the gradients, this method prevents any single loss component from dominating the optimization process. This ensures that the network learns while satisfying all physical constraints concurrently, avoiding poor local minima where the loss is low while the solution is physically inconsistent (e.g., satisfying the boundary conditions perfectly while violating the governing PDE within the domain). This leads to a more stable and effective navigation of the complex optimization landscape.

\subsection{Differentiable Parameterization of Spatial Heterogeneity}
\label{sec:heterogeneity_encoding}

A central innovation of our framework is the differentiable mapping from a low-dimensional parameter vector $\boldsymbol{\lambda}$ to the spatially heterogeneous hydraulic conductivity, represented by the function $K(\mathbf{x}; \boldsymbol{\lambda})$. This mapping must be differentiable with respect to two distinct inputs:
\begin{enumerate}[label=\roman*]
    \item The spatial coordinates $\mathbf{x}$, a feature which is key for computing the PDE residual within the physics-informed loss function.
    \item The parameters $\boldsymbol{\lambda}$, thus turning the trained model into a fully differentiable solver. This property is important for enabling advanced, gradient-based applications such as sensitivity analysis and inverse modeling.
\end{enumerate}
We adapt this general framework to two distinct scenarios, leading to two diverse approximated strategies, as outlined in the following.

\subsubsection{Scenario 1: Functional Parameterization via a Gaussian Anomaly}
\label{sec:functional_param}
In this more direct approach, we define $K$ using a simple analytical function where $\boldsymbol{\lambda}$ controls the geometric properties of a spatial feature. Adoption of this strategy is suitable for scenarios where the nature of the system heterogeneity can be described by a known functional form. In this context, we analyze a system characterized by a constant background conductivity, $K_{\text{min}}$, to which we superimpose a spatial anomaly of circular shape featuring a variation of the value of $K$ and characterized by a maximum conductivity value, $K_{\text{max}}$. The parameter vector $\boldsymbol{\lambda} = (\lambda_1, \lambda_2)$ defines the $(x, y)$ coordinates of the center of this anomaly, allowing it to be placed anywhere within the domain $\Omega$. The spatial distribution of $K(\mathbf{x}$ follow an isotropic Gaussian shape centered around the location $(\lambda_1, \lambda_2)$. The full conductivity field is given by:
\begin{equation}
    K(\mathbf{x}=(x,y); \boldsymbol{\lambda}) = K_{\min} + (K_{\max} - K_{\min}) \exp\left( -\frac{(x - \lambda_1)^2 + (y - \lambda_2)^2}{2 \sigma_K^2} \right),
    \label{eq:k_gaussian_bump_method}
\end{equation}
where $\sigma_K$ is a fixed hyperparameter controlling the spatial extent (width) of the feature. For the numerical analyses performed in this study, the hydraulic conductivity fields are bounded between two fixed values $K_{\text{min}} = \exp(-1.5)$ and $K_{\text{max}} = \exp(1.5)$.

The function defined in eq. \eqref{eq:k_gaussian_bump_method} is differentiable with respect to both $\mathbf{x}$ and $\boldsymbol{\lambda}$, making it straightforward to embed into the computation of the physics-informed loss. In this example the parameter space $\Lambda$ is a subset of the spatial domain $\Omega$, i.e., it identifies the region where the high-conductivity anomaly can be found.

\subsubsection{Scenario 2: Autoencoder-based Latent Space Parameterization}
\label{sec:autoencoder_param}

\begin{figure*}[!ht]
    \centering
    \includegraphics[width=\textwidth]{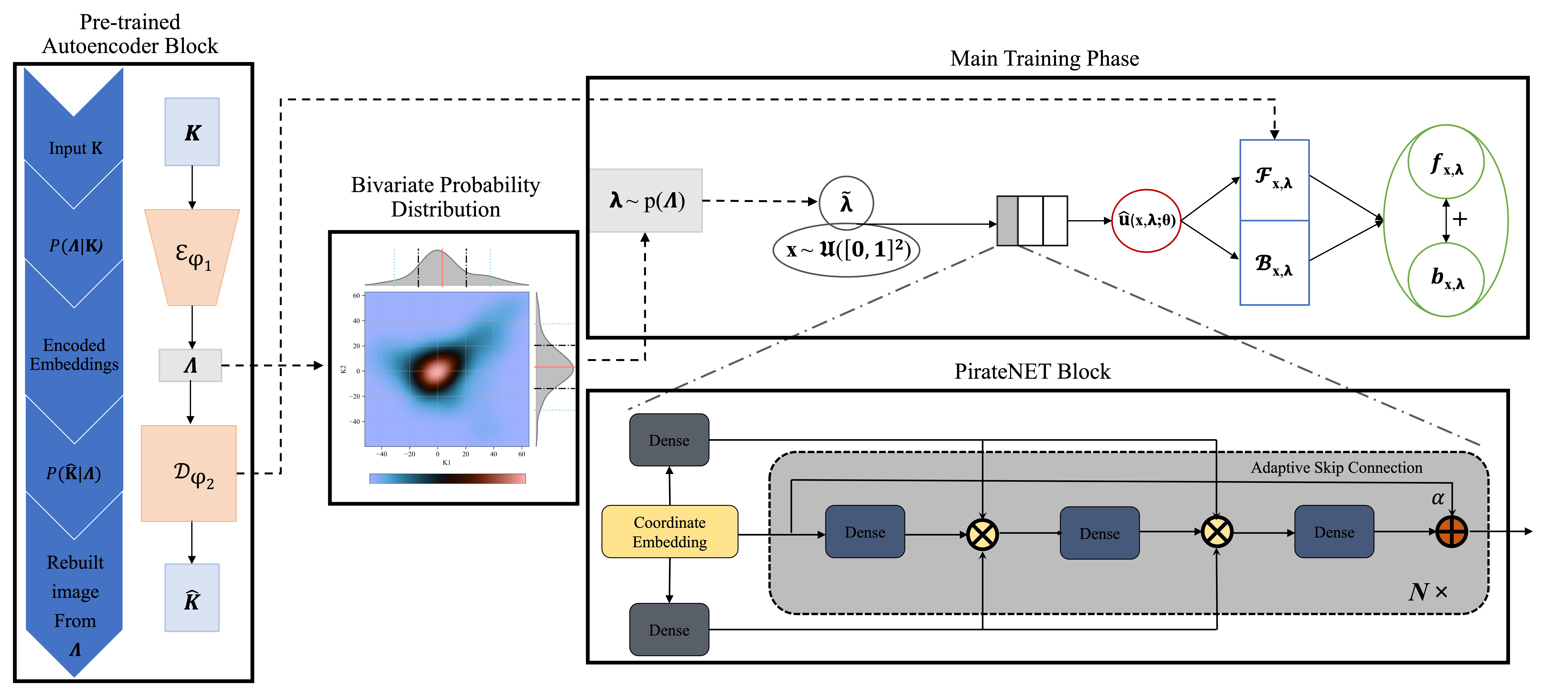} 
    \caption{\textbf{Schematic depiction of the parameterized PINN framework relying on an autoencoder-based hydraulic ($K$) parameterization.} 
    (Left) Autoencoder architecture: a CNN encoder maps reference realizations of $K$ to a 2D latent vector $\boldsymbol{\lambda}$. 
    (Center) Learned two-dimensional (2D) latent space distribution $p(\boldsymbol{\lambda})$, which defines the sampling space for $\boldsymbol{\lambda})$ for PINN training.
    (Top-Right) Parameterized PINN solver, which takes spatial coordinates and latent parameters as input. The differentiable INR decoder $\mathcal{D}_{\phi_2}$ is embedded into the computation of the physics-informed loss. (Bottom-Right) Sketch of the PirateNet architecture (see Supporting Information \ref{app:piratenet}).}
    \label{fig:architecture_autoencoder_case2_method}
\end{figure*}

To generalize the approach to settings where a simple analytical representation of the heterogeneity is not available, we propose a flexible data-driven approach leveraging a pre-trained autoencoder (AE). This corresponding workflow is illustrated in Figure~\ref{fig:architecture_autoencoder_case2_method}. It is designed to enable parameterization of a rich variety of spatially heterogeneous conductivity fields by first learning a low-dimensional latent representation of the heterogeneity and then using this representation to inform the main physics-based solver.

\paragraph{Learning a Generative Model of Heterogeneity.}
The process begins by training an autoencoder (AE) on a dataset of diverse hydraulic conductivity field realizations. These fields are generated via a selected geostatistical workflow, starting from an underlying random field. The specific statistical properties of the reference fields are detailed in the validation Section~\ref{sec:ae_validation} and the full generation algorithm is provided in Supporting Information ~\ref{app:grf_generation}. As shown in Figure~\ref{fig:architecture_autoencoder_case2_method} (pre-trained autoencoder block), the AE consists of a CNN encoder ($\mathcal{E}_{\phi_1}$) that maps each spatial conductivity field to a low-dimensional latent vector, $\boldsymbol{\lambda} \in \mathbb{R}^{d_\lambda}$ (where we use $d_\lambda=2$), and a INR decoder ($\mathcal{D}_{\phi_2}$) that reconstructs conductivity values $\hat{K}$ at any spatial coordinate $\mathbf{x}$.

Once trained on a sample of reference conductivity fields, the collection of encoded vectors $\boldsymbol{\lambda}$ forms an empirical multi-variate distribution, denoted as $p(\boldsymbol{\lambda})$. Since our latent space is two-dimensional ($d_\lambda=2$), the latter is a bivariate probability distribution (as depicted in Figure~\ref{fig:architecture_autoencoder_case2_method}). This distribution defines the sampling space $\Lambda$ that is employed during the training of the main PINN solver. Further architectural and training details for the AE are provided in Supporting Information ~\ref{app:autoencoder}.

\paragraph{Integration of the Differentiable INR Decoder into the PINN.}
In the main training phase (Figure~\ref{fig:architecture_autoencoder_case2_method}; main training phase), the pre-trained INR decoder $\mathcal{D}_{\phi_2}$ is frozen and integrated directly into the PINN's computational graph. The conductivity field is now defined as $K(\mathbf{x}; \boldsymbol{\lambda}) := \mathcal{D}_{\phi_2}(\mathbf{x}, \boldsymbol{\lambda})$. A key innovation of our work is that we leverage the full differentiability of the decoder network. Both the field value $K$ and its spatial derivatives $\nabla K$, which are needed to compute the physics loss, are calculated on-the-fly by applying AD through the decoder. This novel integration enables the PINN to learn the solution manifold across a diverse and complex set of heterogeneous fields, thus demonstrating a powerful synergy between data-driven representation learning and physics-constrained modeling.

\subsection{Training via a Multi-Stage Curriculum}
\label{sec:training_strategy}
Directly optimizing a deep neural network over a high-dimensional and potentially complex domain can present significant challenges, including slow convergence and susceptibility to poor local minima. To address these issues, we employ a multi-stage learning strategy. This approach is a significant extension of the transfer learning concepts successfully applied in our previous work \citep{panahi2025modeling} and aims at stabilizing the training process and accelerate convergence by progressively increasing the complexity of the learning task \citep{Subel2023TL}. The general methodology is formalized in Algorithm~\ref{alg:training_strategy_corr}.

\begin{algorithm}[H]
    \caption{Multi-Stage Curriculum Training for Parameterized PINNs}
    \label{alg:training_strategy_corr}
    \KwData{Governing PDE, network architecture $\mathcal{N}(\cdot;\boldsymbol{\theta})$, total epochs $E$, number of stages $S$}
    \KwResult{Optimized global parameters $\boldsymbol{\theta}_{\text{global}}$}
    \BlankLine
    Initialize network parameters $\boldsymbol{\theta}_0$\;
    \BlankLine
    \tcp*{Initial Stage: Pre-training on the mean parameter field}
    Define parameter subspace $\Lambda_1 \leftarrow \{\bar{\boldsymbol{\lambda}}\}$\;
    Train $\mathcal{N}(\cdot; \boldsymbol{\theta})$ on domain $\Omega \times \Lambda_1$ for $E_1$ epochs to get parameters $\boldsymbol{\theta}_1$\;
    \BlankLine
    \tcp*{Subsequent Stages: Fine-tuning on expanded parameter spaces}
    \For{$s \leftarrow 2$ \KwTo $S$}{
        Define expanded parameter subspace $\Lambda_s$ such that $\Lambda_s \supset \Lambda_{s-1}$\;
        
        Initialize network with weights from $\boldsymbol{\theta}_{s-1}$\;
        
        Train $\mathcal{N}(\cdot; \boldsymbol{\theta})$ on domain $\Omega \times \Lambda_s$ for $E_s$ epochs to get parameters $\boldsymbol{\theta}_s$\;
    }
    \BlankLine
    $\boldsymbol{\theta}_{\text{global}} \leftarrow \boldsymbol{\theta}_S$\;
    \KwRet{$\boldsymbol{\theta}_{\text{global}}$}\;
\end{algorithm}

The core idea underpinning the curriculum described in Algorithm~\ref{alg:training_strategy_corr} is to first allow the network to learn the fundamental solution behavior in a simplified parameter setting before exposing it to the full variability of the parameter space. The efficacy of this strategy can be heuristically grasped through the lens of recent works on phase transitions in PINN training \citep{ANAGNOSTOPOULOS2026107983}. 

Training a PINN from a random initialization often involves a warm-up phase where the optimizer struggles with disordered gradients and heterogeneous residuals. Our multi-stage curriculum is designed to circumvent this inefficient search. The initial pre-training on the mean parameter field guides the network to a state that already captures the low-frequency solution components and establishes a baseline of residual homogeneity. Consequently, when fine-tuning begins on the full parameter space, the network is not starting from a random state, but from a well-conditioned initialization. By progressively expanding the parameter space, our method ensures that the optimizer can efficiently find and maintain a stable equilibrium, leading to faster convergence and a more robust final solution.

\begin{enumerate}[label=\roman*]
    \item \textbf{Stage 1: Pre-training on a Mean or Simplified Field.} Training commences by restricting the parameter vector $\boldsymbol{\lambda}$ to a simplified representation, typically setting all parameters to their mean value, $\bar{\boldsymbol{\lambda}} = \mathbb{E}[\boldsymbol{\lambda}]$. Thus, the initial parameter subspace is $\Lambda_1 = \{\bar{\boldsymbol{\lambda}}\}$. The network $\mathcal{N}(\cdot; \boldsymbol{\theta})$ is trained on the spatial domain $\Omega$. This stage enables the network to learn some average (or effective) physics of the system. Hence, it yields a robust baseline solution.

    \item \textbf{Subsequent Stages: Fine-tuning on Progressively Expanded Parameter Spaces.} After the initial pre-training, the learned network parameters $\boldsymbol{\theta}_1$ serve as a robust and well-conditioned initialization for the next stage. In each subsequent stage $s$ (from $2$ to $S$), we define an expanded parameter subspace $\Lambda_s$ such that $\Lambda_{s-1} \subset \Lambda_s \subseteq \Lambda$. This expansion can be achieved according to the following strategies, depending on the nature of $\boldsymbol{\lambda}$:
        \begin{enumerate}
            \item For the functional parameterization (Gaussian anomaly), if $d_\lambda = 2$, Stage 1 will fix $\boldsymbol{\lambda}$ to the domain center. Stage 2 releases one component (e.g., $\lambda_1$) allowing it to vary while keeping $\lambda_2$ fixed at its mean value. Stage 3 would then release both $\lambda_1$ and $\lambda_2$ to sample from the full two-dimensional parameter space.
            \item For the autoencoder-based parameterization, Stage 1 trains on the mean latent vector $\bar{\boldsymbol{\lambda}}$ (close to $\mathbf{0}$ for a centered latent space). Stage 2 then fine-tunes the network by sampling $\boldsymbol{\lambda}$ from the full learned latent distribution $p(\boldsymbol{\lambda})$.
        \end{enumerate}
    The network is then trained on the collocation points sampled from $\Omega \times \Lambda_s$. This process of initializing with weights from the previous (simpler) stage and fine-tuning on a more complex parameter subspace is repeated until the network is trained on the full target parameter space $\Lambda_S = \Lambda$.
\end{enumerate}
The stage-wise training approach described above acts as a curriculum learning strategy, guiding the optimization process through a sequence of increasingly difficult tasks. By leveraging the knowledge gained in simpler settings, the network is better equipped to navigate the complex loss landscape associated with the full parameterized problem. Doing so leads to more stable training, faster convergence, and (often) more accurate and robust final solutions.

\section{Results and Discussion}
\label{sec:results}
In this section, we illustrate numerical analyses designed to assess the quality of our parameterized PINN framework. We demonstrate its effectiveness and explore its physical consistency across the two distinct heterogeneity encoding strategies detailed in Section~\ref{sec:methodology}, i.e., the functional parameterization via a Gaussian anomaly (section \ref{sec:functional_param}), and the more complex, data-driven approach using a pre-trained autoencoder (section \ref{sec:autoencoder_param}). The problem formulation remains the same for the two test scenarios and is detailed in section \ref{sec:problem_formulation}.

\subsection{Reference Numerical Solutions}
\label{sec:fem_reference}
To quantitatively evaluate the accuracy of our parameterized PINN framework, we compare its outputs against high-fidelity reference solutions generated using a standard Finite Element Method (FEM). The latter provides a robust and well-established numerical solution that is typically considered as baseline for solving groundwater flow settings and is detailed in Supporting Information \ref{sec:appendix_fem}

\begin{figure}[!ht]
    \centering
    \includegraphics[width=.8\textwidth]{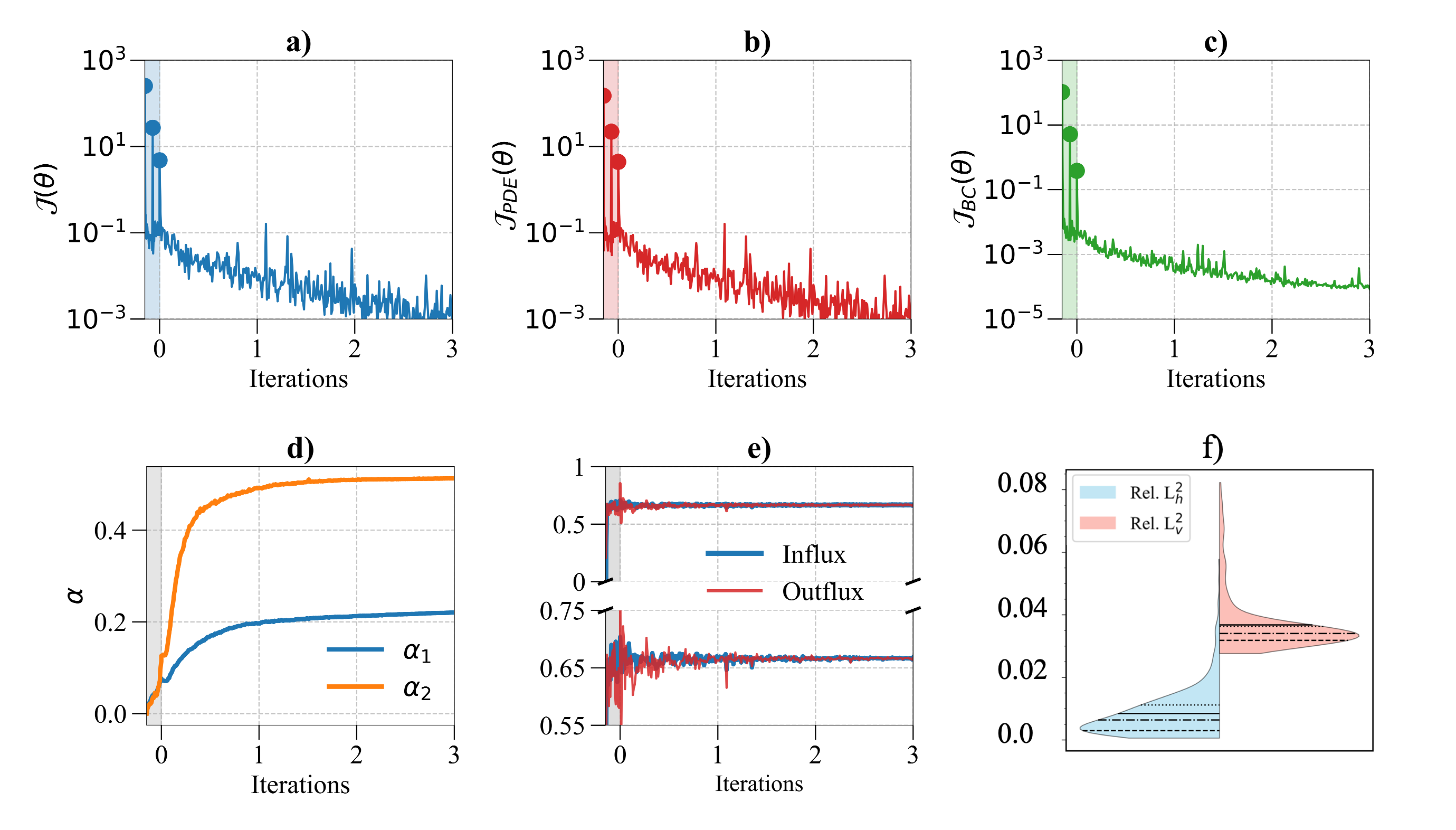} 
    \caption{\textbf{Training dynamics of Scenario 1.} Evolution of (a) the physics-informed loss $\mathcal{J}(\theta)$, (b) the PDE residual component $\mathcal{J}_{PDE}(\theta)$, and (c) the boundary condition residual component $\mathcal{J}_{BC}(\theta)$. (d) Evolution of the trainable $\alpha_i$ ($i = 1, 2$) parameters associated with the adaptive skip connections. (e) Convergence of the mean influx and outflux, demonstrating global mass conservation. Shaded regions in panels (a)-(e) correspond to the initial warm-up (transfer learning) phase where the network is trained on the mean of the latent parameter distribution before training on the full sampled latent space. Filled circles indicate the starting point for each iteration of the training phase (see Section \ref{sec:training_strategy}). (f) Violin plots showing the distribution of $\text{Rel. } L^2_{\hat{h}}$ (light blue) and $\text{Rel. } L^2_{\hat{\mathbf{v}}}$ (salmon) errors over all 1024 $\boldsymbol{\lambda}$ samples mean (solid black line), median (dotted black line), and 25th/75th percentiles (dashed black lines) are also identified.} 
    \label{fig:training_logs_gaussian_case1}
\end{figure}

\begin{figure*}[!ht]
\centering
\includegraphics[width=.8\textwidth]{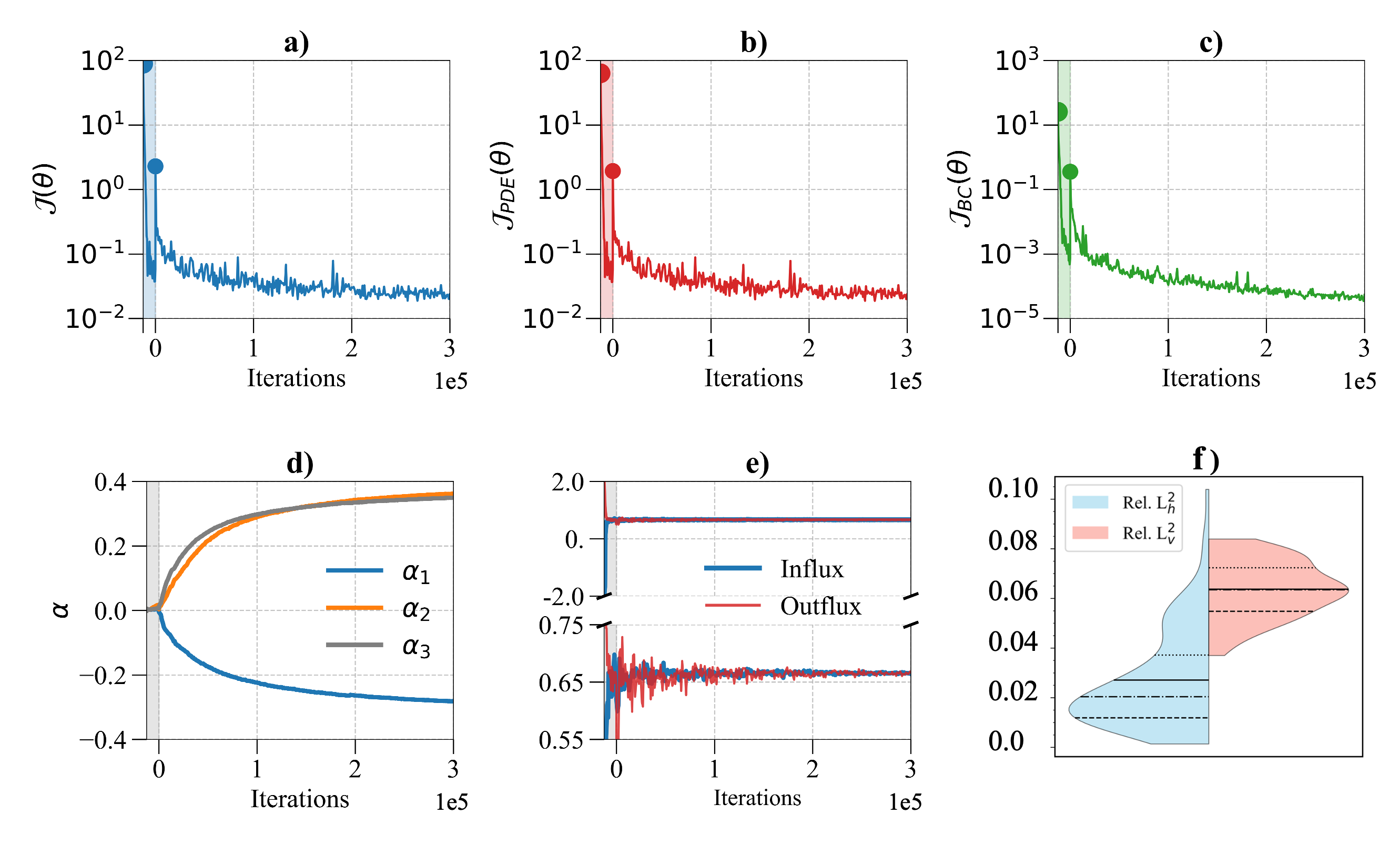}
\caption{\textbf{Training dynamics of Scenario 2.} 
Evolution of (a) the total physics-informed loss $\mathcal{J}(\theta)$, (b) the PDE residual component $\mathcal{J}_{PDE}(\theta)$, and (c) the boundary condition residual component $\mathcal{J}_{BC}(\theta)$. (d) Evolution of the trainable $\alpha_i$ ($i = 1, 2$) parameters associated with the adaptive skip connections.  
(e) Convergence of the mean influx and outflux, the two reported plots display results with different vertical axis scale.
The narrow shaded region in panels (a)-(e) corresponds to the initial warm-up phase where the network is trained on the mean of the latent parameter distribution before training on the full sampled latent space. Filled circles in panels (a)-(c) indicate the starting point for each iteration of the training phase (see section \ref{sec:training_strategy})}
\label{fig:training_logs_autoencoder_case2}.
(f) Violin plots summarizing the distribution of relative $L^2$ errors for computed hydraulic heads ($\text{Rel. } L^2_{\hat{h}}$, light blue) and auto-differentiated velocity ($\text{Rel. } L^2_{\hat{\mathbf{v}}}$, salmon) over 512 test realizations sampled from the latent space, compared to FEM reference solutions. 
\end{figure*}

\subsection{Training Dynamics and Model Convergence}

The performance of the learned (PINN-based) surrogate models is contingent upon a stable and effective training process. Results related to the latter are depicted in Figure~\ref{fig:training_logs_gaussian_case1} and Figure~\ref{fig:training_logs_autoencoder_case2}, for the functional (Scenario 1) and autoencoder-based (Scenario 2) parameterizations, respectively. Note that due to the on-the-fly generation of randomized samples of hydraulic conductivity tensor, all samples are previously unseen, and hence we’ll directly use the training curves here to draw conclusions about the performance of the two approaches.

For the Gaussian anomaly (Figure~\ref{fig:training_logs_gaussian_case1}) the initial warm-up phase, where the model is trained on a single parameter instance corresponding to the center of the parameter range of variability, is highlighted by the shaded region within each subplot. Upon transitioning to training on the full parameter space, the total loss and its constituent components (Figure~\ref{fig:training_logs_gaussian_case1}(a-c)) exhibit a significant initial drop. The latter is then followed by steady convergence. Panel (d) displays the evolution of the trainable parameters $\alpha_i$ ($i = 1, 2$) within the PirateNet adaptive residual blocks (see Supporting Information \ref{app:piratenet}). These parameters are initialized at zero and their value increases during training. As discussed in Supporting Information \ref{app:piratenet}, this trend suggests that the network progressively incorporates non-linearities and increases its effective depth to attain improved fit of the solution manifold. Figure~\ref{fig:training_logs_gaussian_case1}(e) displays the inflow and outflow evaluation, based on the PINN solution. These results confirm that our approach satisfies mass balance with good accuracy. The violin plots (Figure~\ref{fig:training_logs_gaussian_case1}(f)) further summarize these error distributions, showing that the majority of $\boldsymbol{\lambda}$ realizations are associated with low relative errors for both head and velocity. The median relative $L^2$ error related to model-based heads and velocity is approximately 0.01 and 0.035, respectively. This supports the model ability for accurate generalization across the full solution manifold.\\
Results of similar quality in terms of loss function convergence, $\alpha_i$ trends, and global mass balance are also found for the autoencoder-based scenario (see Figure~\ref{fig:training_logs_autoencoder_case2}). Figure~\ref{fig:training_logs_gaussian_case1}(f) summarizes the generalization performance of the fully trained model. The violin plots illustrate the distribution of relative $L_2$ errors for the head and velocity fields across 512 test realizations, supporting the observation that the stability of the training process results in a surrogate with low generalization error.

\begin{figure*}[!ht]
\centering
\includegraphics[width=.8\textwidth]{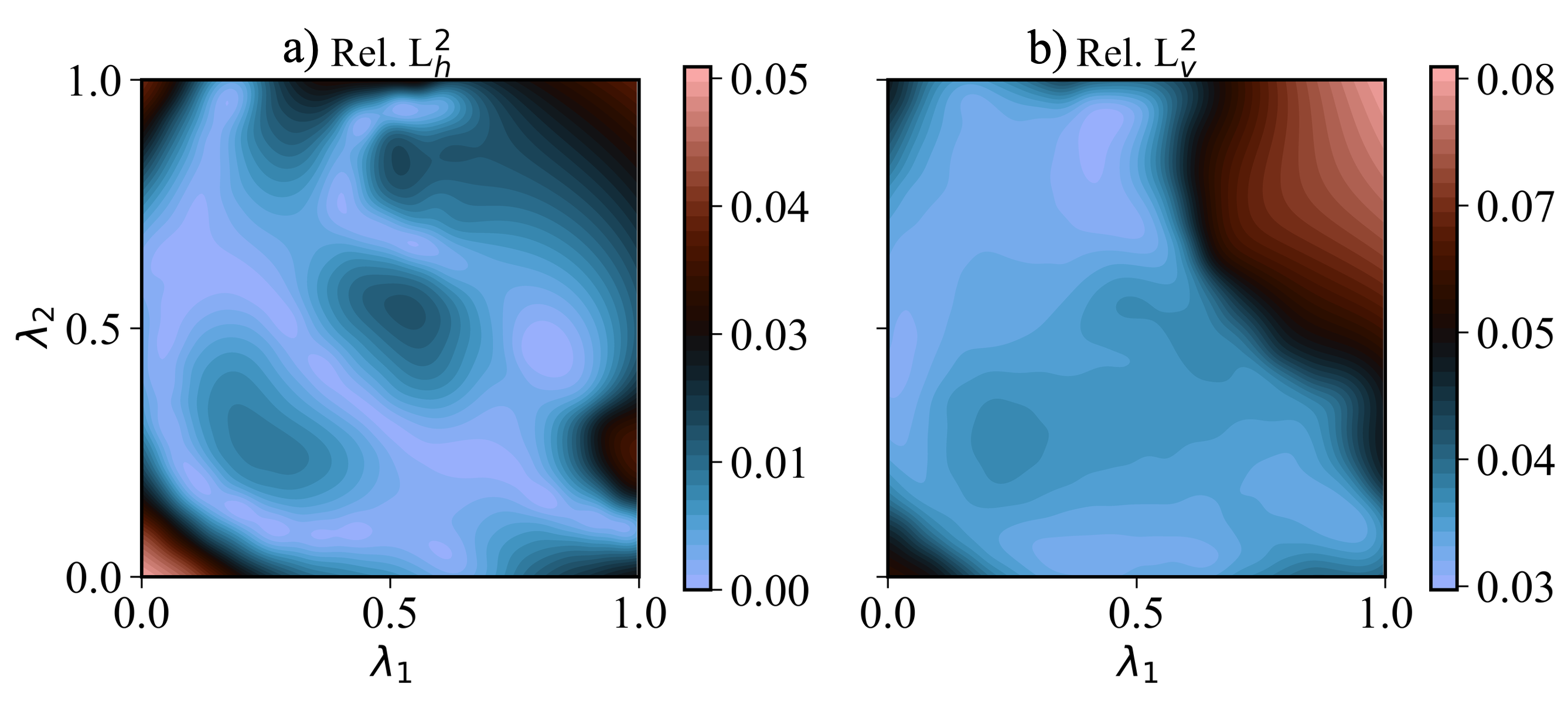}
\caption{\textbf{Performance evaluation of the parameterized PINN trained on the Scenario 1.} 
(a) Contour plot of the relative $L^2$ norm error for model-based hydraulic heads ($\text{Rel. } L^2_{\hat{h}}$) across the $\boldsymbol{\lambda}=(\lambda_1, \lambda_2)$ parameter space, 
(b) Relative $L^2$ norm error for the velocity field ($\text{Rel. } L^2_{\hat{\mathbf{v}}}$).}
\label{fig:error_evaluation_gaussian_case1}
\end{figure*}

\subsection{Scenario 1: Functional Parameterization}
We first evaluate the framework on the simple case where hydraulic conductivity is defined by a Gaussian anomaly whose center is controlled by the parameter vector $\boldsymbol{\lambda} = (\lambda_1, \lambda_2)$.
\paragraph{Global Accuracy and Generalization.}
Figure~\ref{fig:error_evaluation_gaussian_case1} depicts the results of a quantitative assessment of the model accuracy, evaluated over a grid of 1024 distinct $\boldsymbol{\lambda}$ values and compared against reference solutions from the Finite Element Method solver. The contour plots (Figure~\ref{fig:error_evaluation_gaussian_case1}(a, b)) illustrate the spatial distribution of the relative $L_2$ error associated with the model-based hydraulic head ($\hat{h}$) and the derived velocity field ($\hat{\mathbf{v}}$). The relative $L_2$ error for the velocity field is computed considering both components, thereby accounting for discrepancies in both magnitude and direction. The framework demonstrates robust generalization, with low error across the entire parameter space. Errors in the computed velocity markedly increase as the Gaussian anomaly interferes with the outflow boundary conditions (see top-right corner in Figure~\ref{fig:error_evaluation_gaussian_case1}(b)). This result is related to the presence of sharp velocity gradients close to the outlet section, that are challenging to approximate in the presence of local variations of $K$.

\begin{figure}[!ht]
    \centering
    \renewcommand{\thesubfigure}{\roman{subfigure}}

    \begin{subfigure}{0.85\textwidth}
        \centering
        \includegraphics[width=\textwidth]{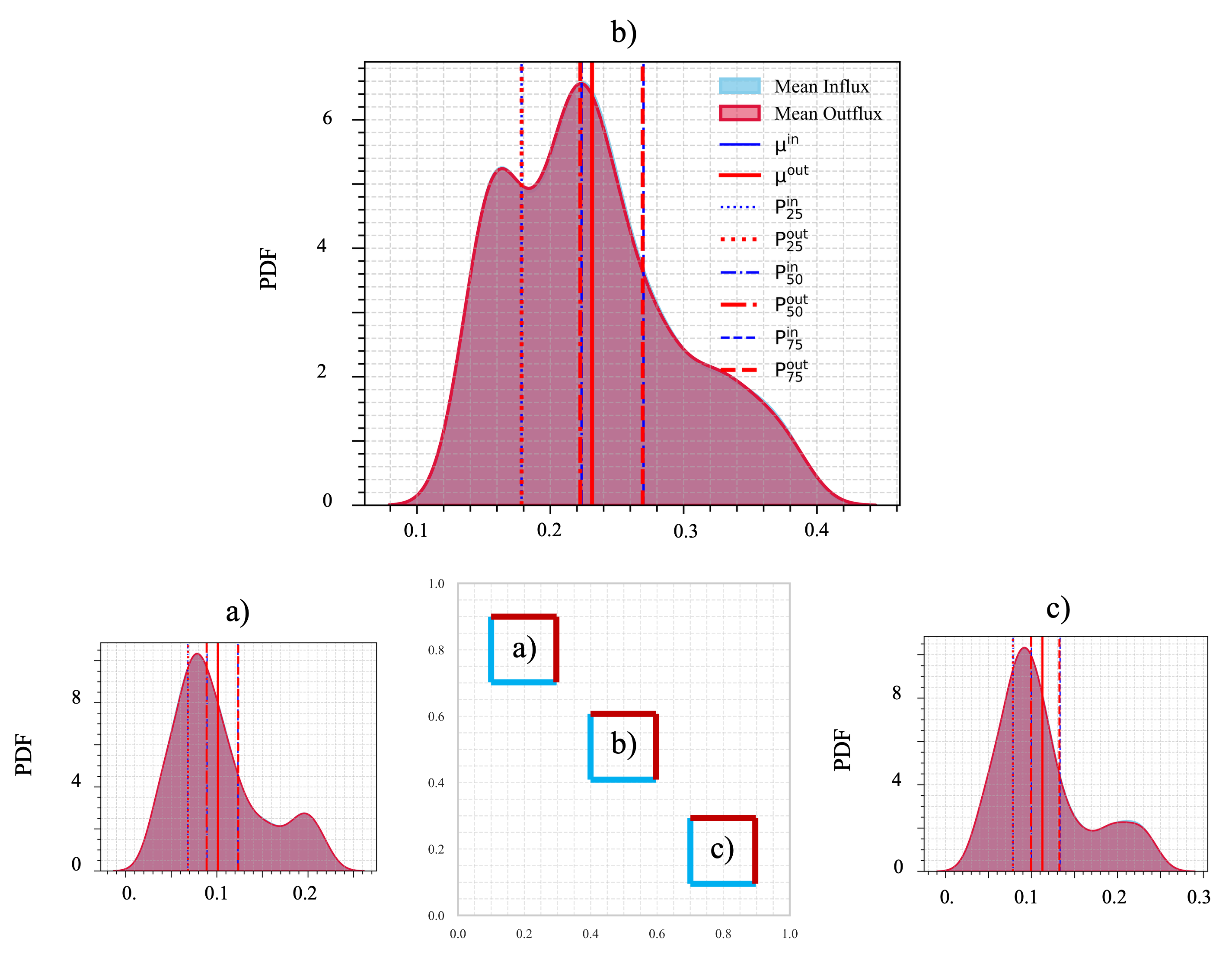}
        \caption{Comparison of the probability density functions (PDFs) of total influx and outflux for the three control volumes depicted in the central panel.}
        \label{fig:local_conservation_gaussian_case1}
    \end{subfigure}

    \vspace{1em} 

    \begin{subfigure}{0.85\textwidth}
        \centering
        \includegraphics[width=\textwidth]{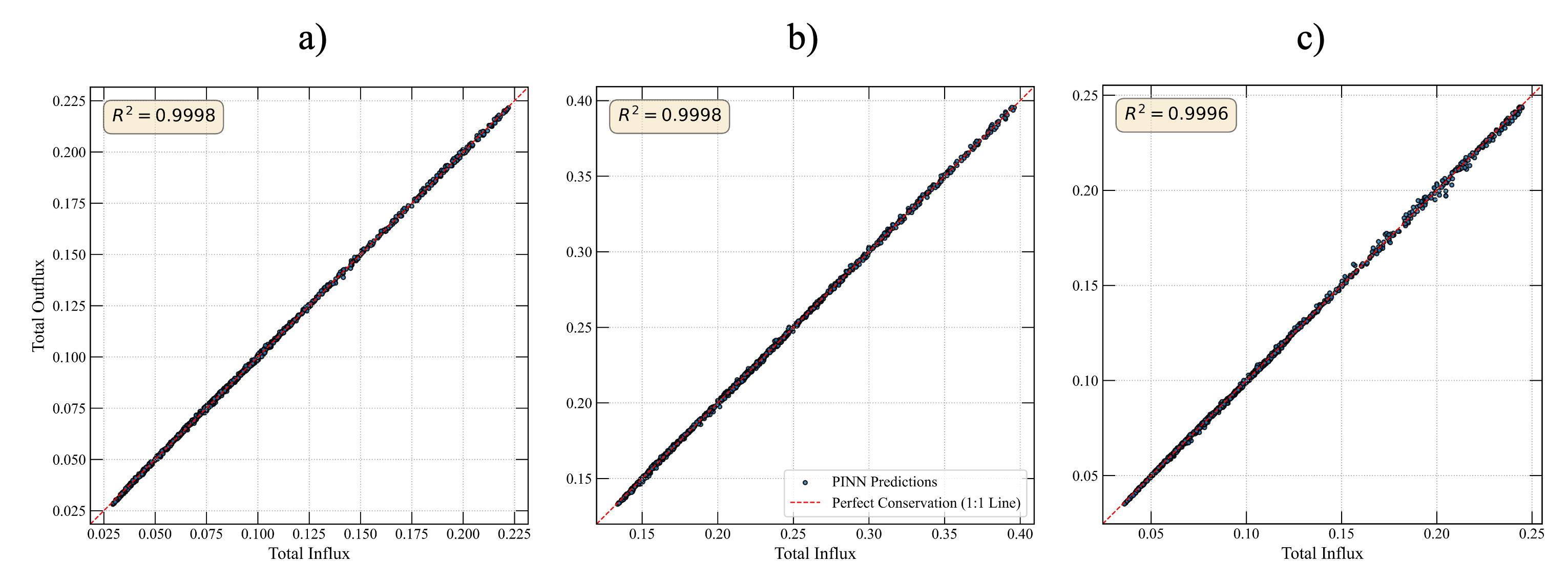}
        \caption{Pairwise scatter plots of outflux versus influx for each realization in the ensemble, corresponding to the three control volumes.}
        \label{fig:local_conservation_scatter}
    \end{subfigure}
    
    \caption{\textbf{Assessment of local mass conservation for the Scenario 1.} 
    Two complementary analyses are displayed: \textbf{(i)} comparison of the statistics via the PDFs of total influx and outflux for the three control volumes considered (a-c), \textbf{(ii)} pairwise validation via scatter plots for each realization. The tight clustering of all points along the 1:1 line of perfect conservation ($R^2 > 0.999$ in all cases) quantitatively supports attainment of local mass conservation.}
    \label{fig:local_conservation_combined}
\end{figure}

\begin{figure*}[!ht]
\centering
\includegraphics[width=.6\textwidth]{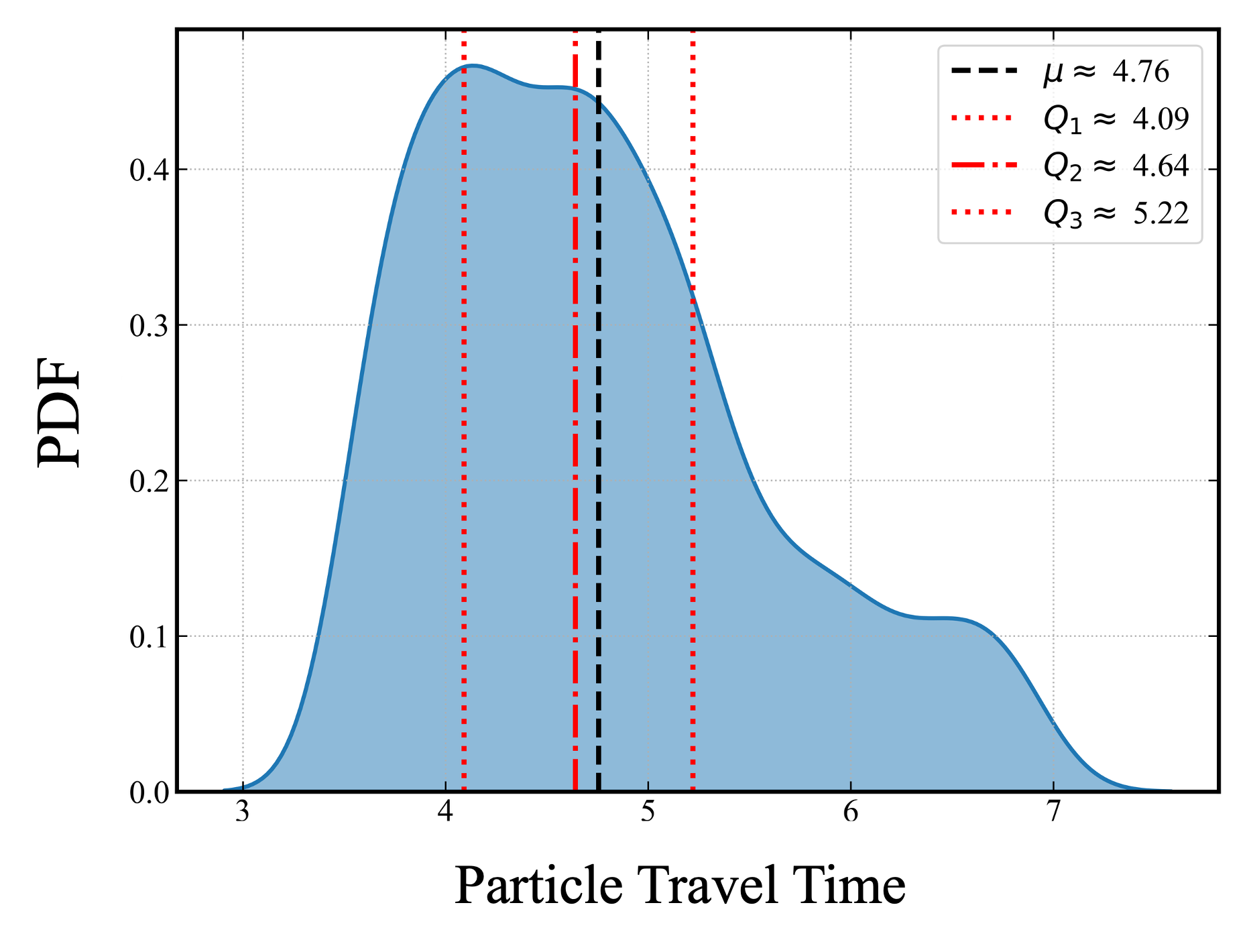}
\caption{\textbf{Travel time distribution for the Gaussian anomaly test case.} 
Probability density function (PDF; estimated via Kernel Density Estimation (KDE)) of inlet-to-outlet travel times estimated via advective particle tracking, considering injection of a particle at location $x_0=0.1$ m, $y_0=0.001$ m).
The sample mean (dashed black line, 4.36 d) and the quartiles $Q_i$ of the distribution are denoted through vertical lines (dotted red line, 4.19 d).}
\label{fig:particle_travel_time_distribution_gaussian_case1} 
\end{figure*}

\paragraph{Local Mass Conservation.}
Beyond global error metrics, we investigate whether the learned solutions adhere to fundamental physical principles at a local level. Satisfaction of local mass conservation is a key indicator of a physically consistent solution. We analyze this feature upon considering three selected control volumes within the domain (labeled a-c in Figure~\ref{fig:local_conservation_gaussian_case1}).

As a first analysis (Figure~\ref{fig:local_conservation_gaussian_case1}), we compare the probability density functions (PDFs) of the total influx and outflux for each control volume across the 1024-realization ensemble. The near-perfect overlap of the distributions suggests that influx and outflux are equivalent from a statistical standpoint. We further perform a direct, pairwise comparison of fluxes (Figure~\ref{fig:local_conservation_scatter}). Each point in these scatter plots represents a single parameter realization, values of computed influx and outflux being associated with horizontal and vertical axis, respectively. These results demonstrate a remarkable level of achievement of local mass conservation. Data points align along the 1:1 line, which represents perfect conservation (i.e., $influx = outflux$), for all three control volumes. These results imbue us with confidence that the PINN, despite being trained only on the global PDE residual, successfully learns solutions that implicitly and robustly satisfy the fundamental principle of local mass conservation on a per-realization basis.

\begin{figure*}[!ht]
    \centering
    \includegraphics[width=1.\textwidth]{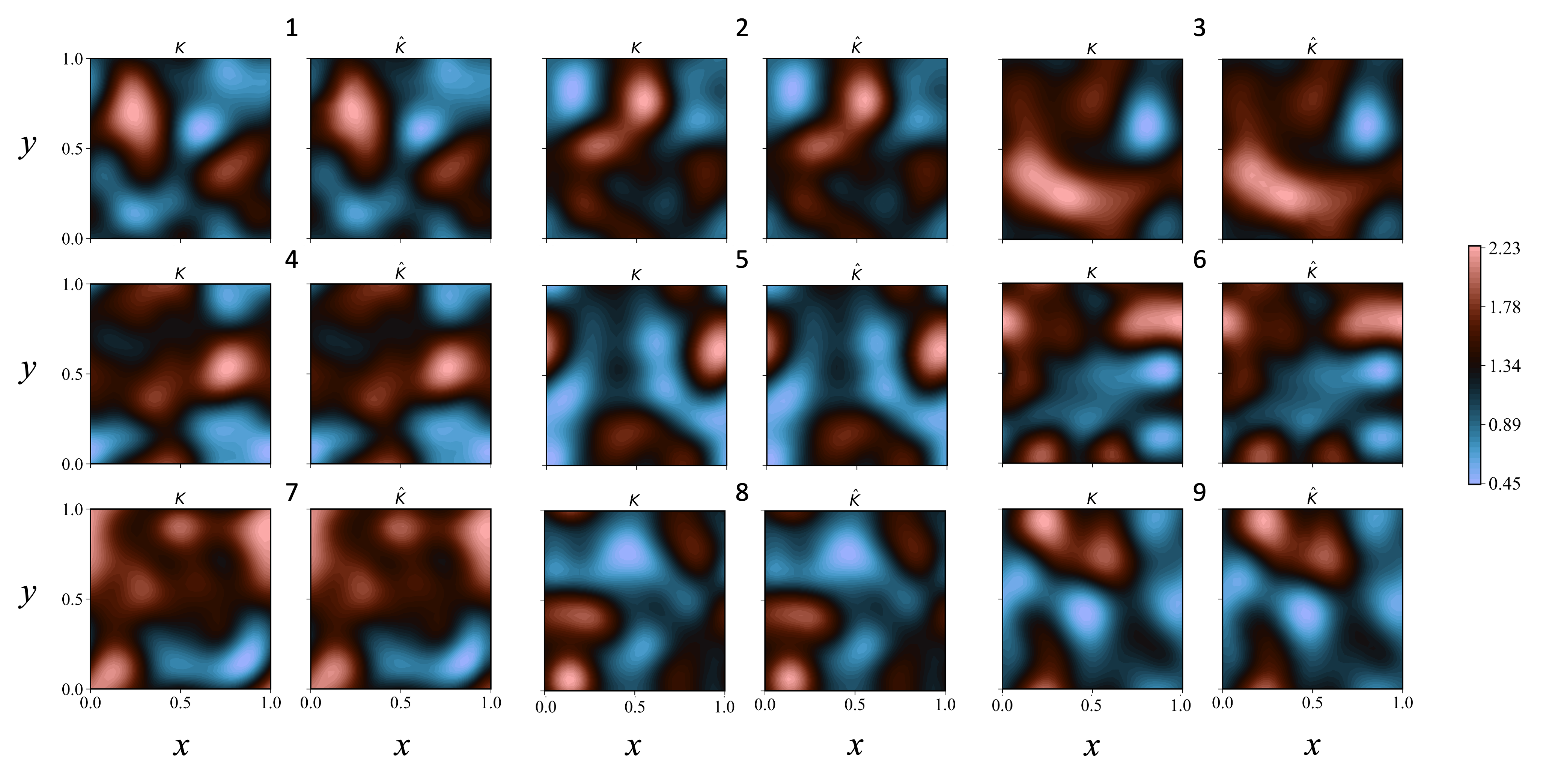}
    \caption{\textbf{Qualitative assessment of the autoencoder reconstruction capability.}
    Nine representative pairs of original conductivity fields ($K$) from the training dataset and their corresponding reconstructions ($\hat{K}$) by the trained INR decoder. Each numbered pair displays the ground truth field alongside the field reconstructed by the decoder from its 2D latent representation. The color bar indicates the magnitude of the normalized hydraulic conductivity tensor values. The high degree of visual similarity confirms the decoder ability to generate complex heterogeneous fields.}
    \label{fig:ae_reconstruction_samples_case2}
\end{figure*}

\begin{figure}[!ht]
    \centering
    \includegraphics[width=.8\textwidth]{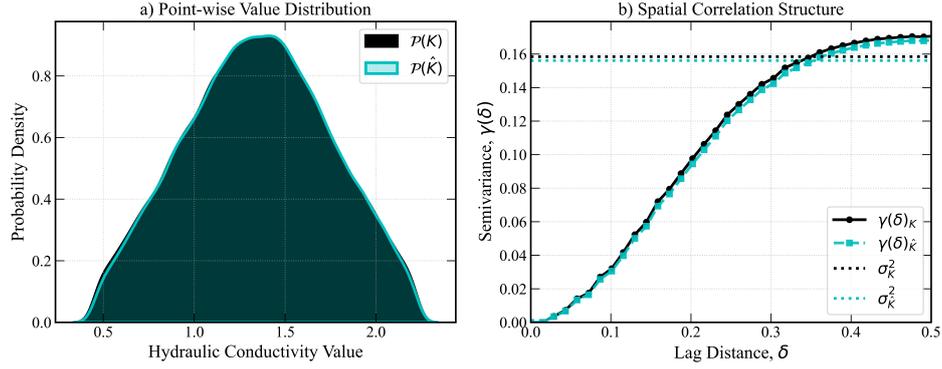}
    \caption{\textbf{Quantitative assessment of the autoencoder reconstruction capability.} The plots compare statistical properties of an ensemble of original training $K$ fields and their corresponding (model-based) reconstructions $\hat{K}$. \textbf{(a)} PDFs of $K$ and $\hat{K}$; \textbf{(b)} comparison of empirical semivariograms for the two fields.} 
    \label{fig:ae_statistical_validation}
\end{figure}

\begin{figure*}[!ht]
\centering
\includegraphics[width=.9\textwidth]{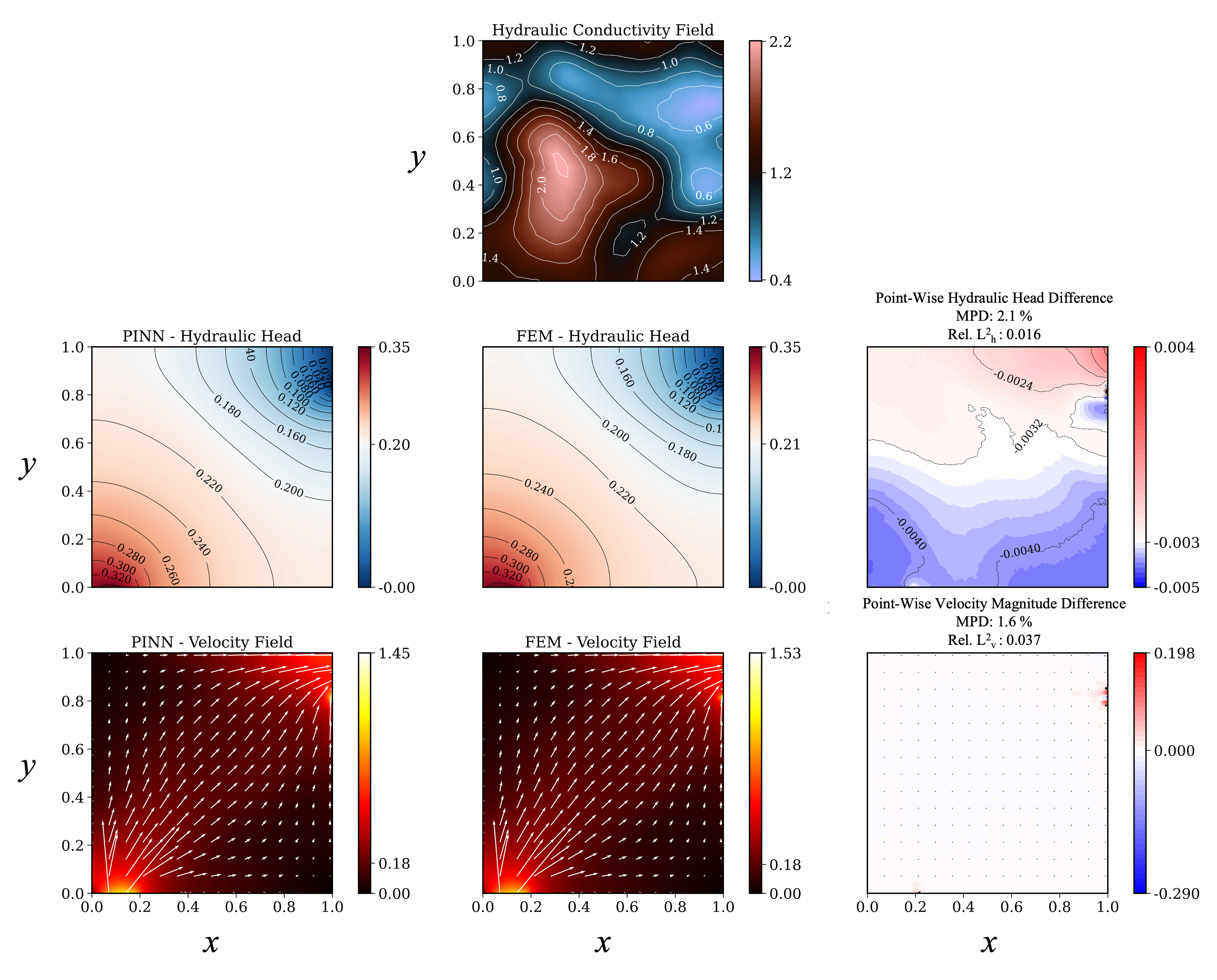} 
\caption{\textbf{Exemplary PINN solution for a hydraulic conductivity from the autoencoder latent space, corresponding to the best-case scenario (lowest relative $L^2$ error in velocity) among 512 test realizations.} 
(Top Row) The specific heterogeneous hydraulic conductivity $K(\mathbf{x}; \boldsymbol{\lambda}_{best})$ generated by the INR decoder for the selected latent vector $\boldsymbol{\lambda}_best)$.
(Middle Row, from left to right) Hydraulic head field resulting from the PINN ($\hat{h}$); reference hydraulic head field obtained through the FEM solver ($h_{\text{FEM}}$); point-wise difference ($h_{\text{FEM}} - \hat{h}$) between FEM and PINN head solutions (Mean Percentage Difference (MPD) and relative $L^2$ norm error are also indicated).
(Bottom Row, from left to right) Velocity field (magnitude contours and quiver plot)  ($\hat{\mathbf{v}}$); reference velocity field from FEM solver ($\mathbf{v}_{\text{FEM}}$); point-wise difference in velocity magnitude ($\|\mathbf{v}_{\text{FEM}}\| - \|\hat{\mathbf{v}}\|$) between FEM- and PINN-based solutions, including MPD and relative $L^2$ norm error for velocity magnitude.}
\label{fig:best_case_autoencoder_case2} 
\end{figure*}

\begin{figure*}[!ht]
\centering
\includegraphics[width=.9\textwidth]{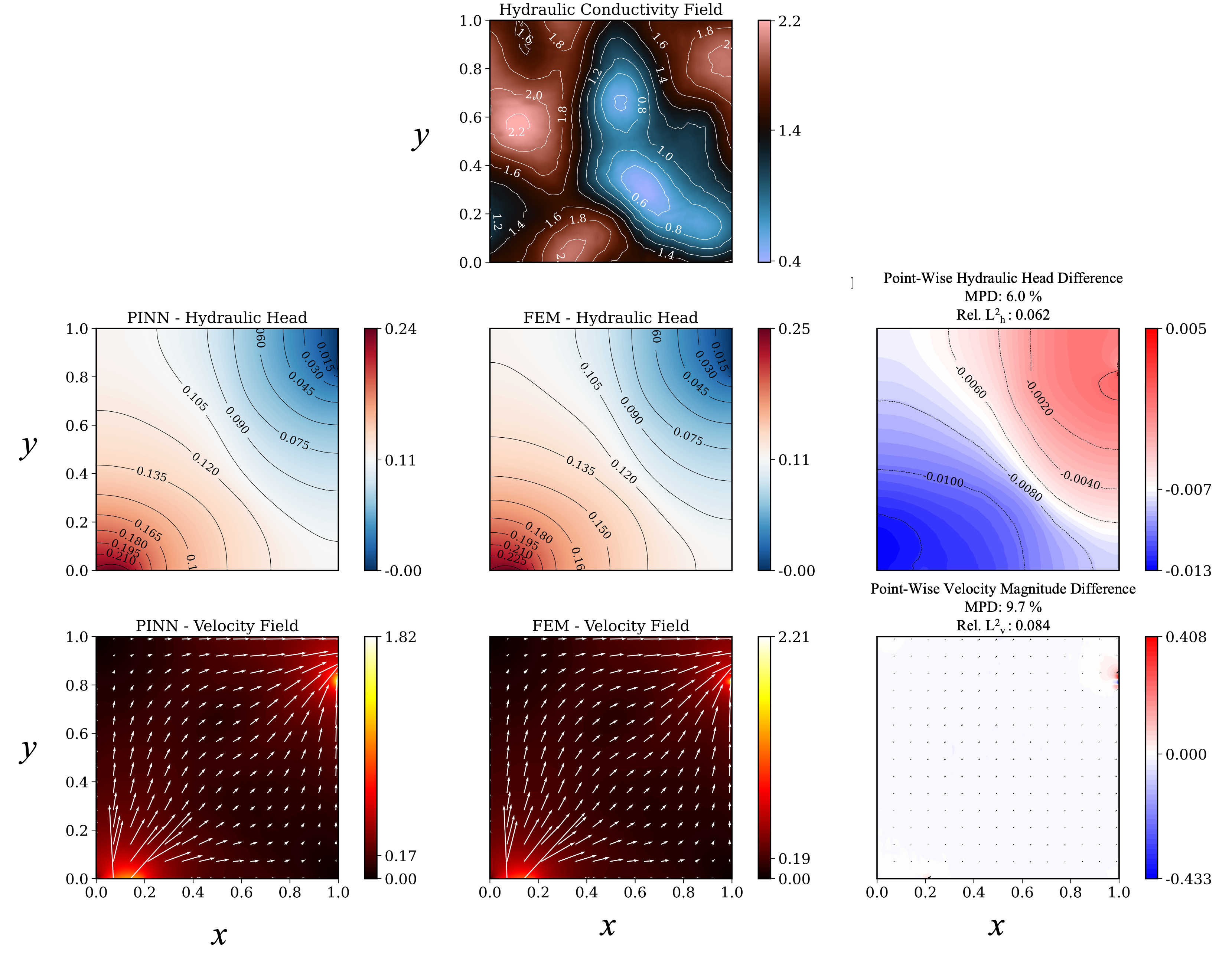} 
\caption{\textbf{Exemplary PINN solution for a hydraulic conductivity from the autoencoder latent space, corresponding to the worst-case scenario (highest relative $L^2$ error in velocity) among 512 test realizations.}
(Top Row) The specific heterogeneous hydraulic conductivity field $K(\mathbf{x}; \boldsymbol{\lambda}_{worst})$ generated by the INR decoder for the selected latent vector $\boldsymbol{\lambda}_{worst}$.
(Middle Row, from left to right) Hydraulic head field evaluated through the PINN ($\hat{h}$); reference hydraulic head field based on the FEM solver ($h_{\text{FEM}}$); point-wise difference ($h_{\text{FEM}} - \hat{h}$) between FEM and PINN head solutions (Mean Percentage Difference (MPD) and relative $L^2$ norm error are also indicated).
(Bottom Row, from left to right) Velocity field (magnitude contours and quiver plot) derived from the PINN's head output via automatic differentiation ($\hat{\mathbf{v}}$); reference velocity field from FEM solver ($\mathbf{v}_{\text{FEM}}$); point-wise difference in velocity magnitude ($\|\mathbf{v}_{\text{FEM}}\| - \|\hat{\mathbf{v}}\|$) between FEM- and PINN-based solutions, including MPD and relative $L^2$ norm error for velocity magnitude.}
\label{fig:worst_case_autoencoder_case2} 
\end{figure*}

\begin{figure*}[!ht]
\centering
\includegraphics[width=.8\textwidth]{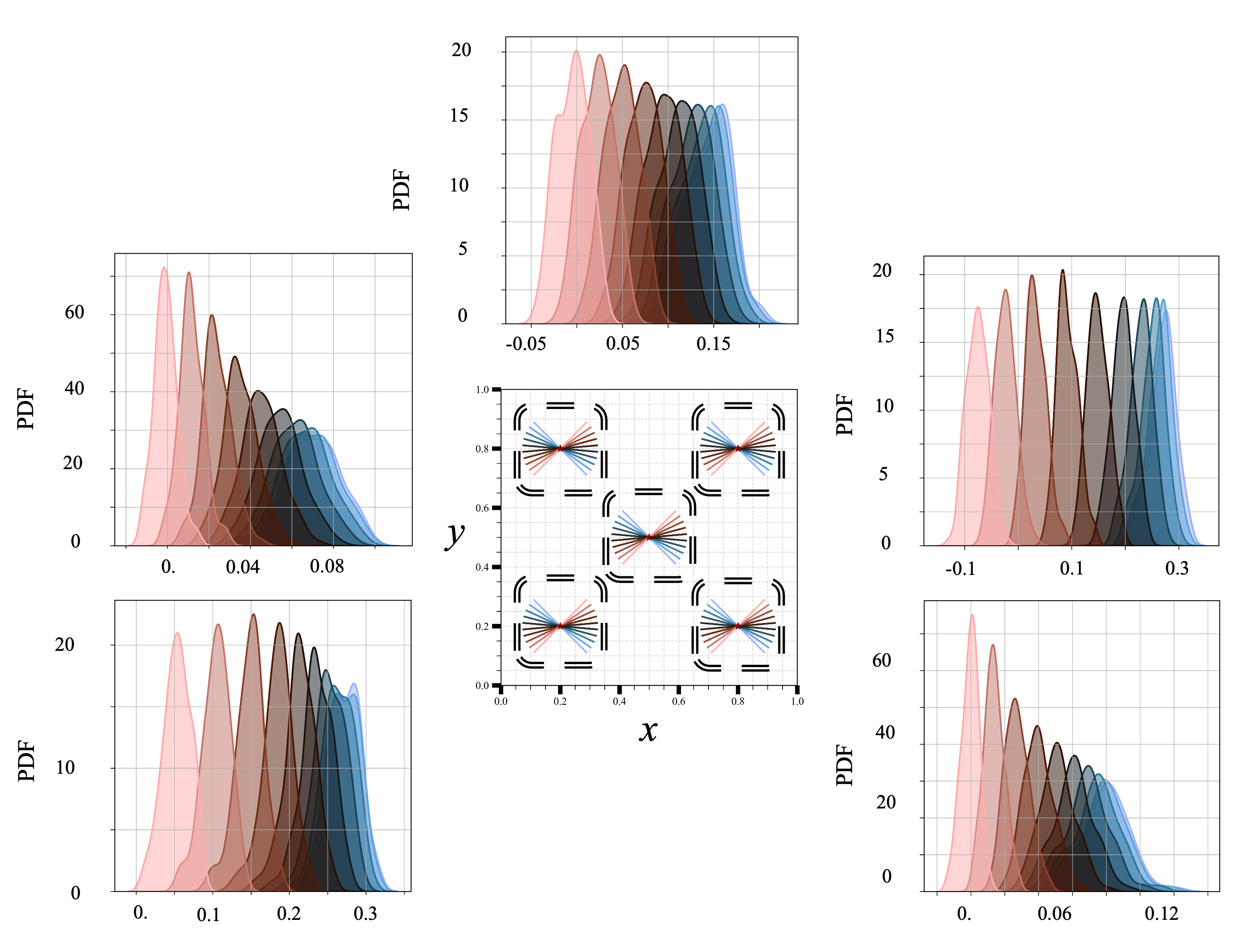} 
\caption{\textbf{Results of the analysis of directional flux variability stemming from the parameterized PINN with autoencoder-based conductivity.} 
The central panel depicts five distinct locations within the unit domain where a series of short line segments (0.2 m) are oriented at different angles (e.g., increments of $22.5^\circ$ from $0^\circ$ to $157.5^\circ$, totaling 8 orientations). 
For each oriented line segment, the sample probability density function (PDF) of the net flux (component of velocity normal to the line, integrated along its length) is assessed across all 512 test realizations of $\boldsymbol{\lambda}$ (conductivity fields). 
Surrounding subplots display these PDFs for their corresponding central point and line orientations. Each PDF is color-coded according to the corresponding line angle. 
The spread and modal shifts of these PDFs serve as indicators to illustrate the anisotropic nature of the overall uncertainty associated with flux magnitude and direction as captured by the our model across diverse heterogeneous fields.}
\label{fig:directional_flux_variability_autoencoder_case2} 
\end{figure*}

\begin{figure}[!ht]
    \centering
    \includegraphics[width=0.5\textwidth]{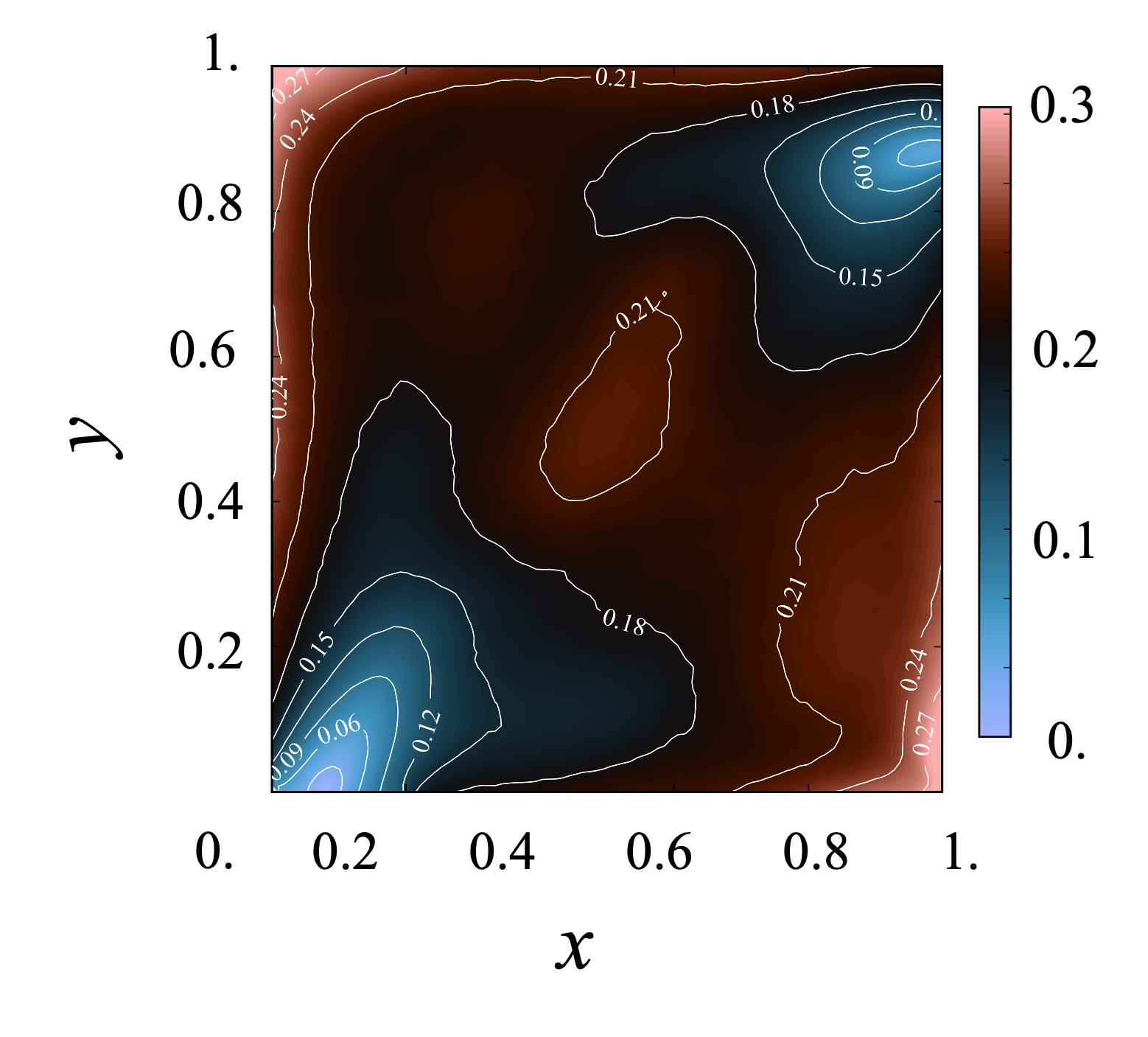}
    \caption{\textbf{Spatial map of relative uncertainty associated with the flow field.} Coefficient of Variation (CV) of the velocity norm, evaluated across an ensemble of 512 conductivity realizations.}
    \label{fig:velocity_cv_map_contours}
\end{figure}

\paragraph{Application to Uncertainty Quantification.}
To showcase a practical application of the framework, we perform a Monte Carlo analysis considering a set of particle tracking simulations to estimate the distribution of advective travel times. The results of this analysis are depicted in Figure~\ref{fig:particle_travel_time_distribution_gaussian_case1}. The Monte Carlo ensemble is formed by $2^{10}=1024$ distinct hydraulic conductivity realizations randomly generated by sampling from the parameter space $\Lambda$.

For each of these 1024 realizations, we simulate the trajectory of a single particle. The latter is introduced at a fixed starting location at the domain inlet, $\mathbf{x}_0 = (0.1, 0.001)\,\text{m}$, and is advected across the corresponding steady-state velocity field stemming from the trained PINN surrogate. The particle movement is computed using a first-order Forward Euler integration scheme with a fixed time step of $\Delta t = 0.01\,\text{d}$. 

A given simulation is terminated once the particle reaches the defined outlet region ($\Gamma_D$), the total elapsed time is recorded as its advective travel time. The resulting distribution of these 1024 travel times reveals significant variability (with a mean of 4.76\,d and a median of 4.64\,d). Our results display a skewed distribution, being characterized by a mild right-tail. This outcome is likely due to the slight asymmetry of the considered setup, resulting from the imposed inlet-outlet location, which in turn results in an asymmetric effect of the hydraulic conductivity perturbation.

\subsection{Scenario 2: Autoencoder-based Latent Space Parameterization}

Here, we focus on the more complex case where heterogeneity is defined by a low-dimensional latent vector $\boldsymbol{\lambda}$ from a pre-trained autoencoder.

\subsubsection{Validation of the Autoencoder-Based Parameterization}
\label{sec:ae_validation}
To validate the autoencoder, we must first define the reference hydraulic conductivity $K$, it was trained to reproduce. The generation process (detailed in Supporting information ~\ref{app:grf_generation}) starts by creating an ensemble of standard, zero-mean Gaussian Random Fields (GRFs) and an isotropic correlation length of $\ell = 0.15$. In the hydrogeological context, this underlying field is conventionally interpreted as the log-conductivity, $Y = \ln(K)$. These raw $Y$  fields are then subjected to a deterministic transformation (exponentiation followed by linear scaling, see Supporting information ~\ref{app:data_preprocessing}) to produce the final training dataset with $K \in [e^{-0.8}, e^{0.8}] \approx [0.45, 2.22]$.

The latter transformation defines the statistical properties of the reference ensemble used for our analysis. The fields are characterized by an ensemble mean conductivity of $\bar{K} \approx 1.35$ and a total ensemble variance of $\approx 0.1583$ (represented by the black dotted line in Fig.~\ref{fig:ae_statistical_validation} b). We note that this total variance is the sum of the average spatial variance within realizations $\approx 0.146$ and the variance of the means across realizations $\approx 0.0123$. By definition, for a stationary field, the sill of the semivariogram (corresponding to the plateau attained at large distances where data are uncorrelated) equals the variance of the field. As observed in the spatial correlation structure, the semivariogram sill stabilizes at a slightly higher value ($\gamma_{sill} \approx 0.17$). This discrepancy is a well-known statistical consequence of the finite domain size relative to the correlation length ($L \approx 6.6\ell$). In finite domains, the sample variance tends to underestimate the true global variance of the process because it is calculated relative to the local mean of the realizations rather than the global population mean. The variogram sill, therefore, provides a more robust estimate of the total structural variability (ergodicity) of the generated fields. Overall, this analysis provides evidence that our autoencoder successfully preserves second-order spatial statistics, including both the field correlation length and the total spatial variance ($\sigma_K^2$).

\paragraph{Assessment of the autoencoder performance}
An initial, qualitative assessment of the autoencoder performance is provided in Figure~\ref{fig:ae_reconstruction_samples_case2}. The latter displays nine pairs of reference  conductivity fields alongside their corresponding reconstructions obtained through the INR decoder from their 2D latent representations. The approximated fields display a striking similarity to the reference ones also in terms of fine-scale details, supports the ability of the autoencoder to successfully learn a viable compression of the salient features of the heterogeneous media.

To further explore the potential of the encoder, we perform a quantitative analysis grounded on the comparison between two key statistical features across an ensemble of 512 fields. Figure~\ref{fig:ae_statistical_validation} embeds the results of this analysis. We start by comparing the point-wise PDF of the conductivity values (Figure~\ref{fig:ae_statistical_validation} a). The overlap of the distributions for the original ($\mathcal{P}(K)$) and reconstructed ($\mathcal{P}(\hat{K})$) fields provides strong evidence that the autoencoder accurately preserves the statistics (the correct range and frequency distribution of conductivity values) of the reference field. We then compare the spatial correlation structure by evaluating the empirical semivariograms (Figure~\ref{fig:ae_statistical_validation} b). We observe a satisfactory agreement between the semivariograms of the original ($\gamma(\delta)_K$) and reconstructed ($\gamma(\delta)_{\hat{K}}$) ensembles.

Note that the quality of the performance evidenced here is related to the specific characteristics of the considered reference fields. A comprehensive analysis across heterogeneous fields displaying a wide range of statistical properties and correlation structures is beyond the scope of this study.

\paragraph{Performance on Individual Realizations.}
To illustrate the performance in specific instances, Figure~\ref{fig:best_case_autoencoder_case2} and Figure~\ref{fig:worst_case_autoencoder_case2} depict exemplary results corresponding to the best- ($K(\mathbf{x}; \boldsymbol{\lambda}_{best})$) and worst-case ($K(\mathbf{x}; \boldsymbol{\lambda}_{worst})$) scenarios (in terms of velocity error, evaluated against the FEM reference solution). Results are evaluated from a sample of 512 Monte Carlo realizations. In the best-case setting (Figure~\ref{fig:best_case_autoencoder_case2}), PINN results for both head and velocity display excellent agreement with their FEM reference counterparts, with a Mean Percentage Difference (MPD) of 2.1\% for head and 1.6\% for velocity. In the worst-case scenario (Figure~\ref{fig:worst_case_autoencoder_case2}), the PINN still maintains a good qualitative agreement with the reference solution, with an MPD of 6.0\% for head and 9.7\% for velocity. Notably, this particular realization is characterized by the presence of a high conductivity region close to the outlet. This feature could underpin the relatively large errors obtained across that particular region. While representing an outlier, this setting helps to define the performance envelope of the model and highlights its robustness even for complex conductivity structures.

\paragraph{Analysis of Flow Characteristics.}
To explore the directional characteristics and inherent uncertainty of the flow field predicted by our parameterized PINN, we analyze the flux across line segments at varying orientations. Figure~\ref{fig:directional_flux_variability_autoencoder_case2} presents this analysis for the autoencoder-based case. At five distinct central points within the domain, a set of co-located short line segments (length 0.2 m) were defined, each rotated to a different orientation.

For each of the 512 conductivity realizations, the net flux through each oriented line segment is calculated by integrating the PINN-derived velocity component normal to the line. The resulting sample PDFs of these directional fluxes are displayed in the surrounding subplots, each associated with its corresponding line orientation and central position. The observed variations in the shape, spread, and central tendency (mean or modal value) of these PDFs across different angles and spatial locations underlines the anisotropic nature of the flux uncertainty. For instance, at a given location, some orientations yield narrowly distributed flux values, while others exhibit broader variability, reflecting the combined effects of heterogeneity and boundary conditions. Our approach enables one to efficiently perform this type of analysis. The results of the latter reveal preferential flow directions and flux sensitivity to conductivity variations. We recall that these results would be prohibitively expensive to obtain via conventional Monte Carlo simulations, especially when evaluating fluxes along arbitrary lines across the domain.
The PINN ability to provide continuous, differentiable solution fields allows for such flexible post-processing and interrogation of the flow field.

Finally, we characterize the spatial distribution of flow uncertainty. We recall (see Section~\ref{sec:ae_validation}) that the input heterogeneity is driven by an underlying conductivity field, $K$, with a variance of $\sigma^2_K \approx 0.1583$. This value, representing the $input uncertainty$, corresponds to a moderate level of heterogeneity, characteristic of many sedimentary aquifer systems. In the following we analyze the resulting $output uncertainty$ upon relying on the local value of the Coefficient of Variation ($\text{CV} = \sigma / \mu$) of the velocity norm.

Figure~\ref{fig:velocity_cv_map_contours} presents the spatial map of CV values of velocity norms. Near the inlet and outlet, the flow is strongly constrained by the prescribed boundary conditions, leading to low velocity variability (and thus low CV), as observed in the plot. In contrast, the flow paths in the interior of the domain are highly sensitive to the specific realization of the heterogeneous conductivity field. It is in these regions, where the flow must navigate the complex conductivity field, that the velocity exhibits the highest variability relative to its mean, leading to the high CV values. This quantitative view of relative uncertainty, which highlights the areas of greatest predictive uncertainty away from the boundaries, offers comprehensive insights into the spatial structure of the flow field statistics.

\section{Conclusions and Future Perspectives}
\label{sec:conclusion}

This study introduces a general and scalable framework recasting a Physics-Informed Neural Network (PINN) as a differentiable physics solver. The latter is exemplified through the solution of steady-state Darcy flow taking place in heterogeneous porous media. As a distinctive feature, the proposed approach captures the effects of parameter variability, thereby addressing the pervasive uncertainty arising from incomplete knowledge of subsurface properties. By training a single, unified network on physics-based residuals over a joint spatio-parameter domain, the model learns an entire manifold of physically consistent solutions. It then enables rapid, flexible, and accurate inference for any parameter instance. This work demonstrates how differentiable physics-based learning can bridge traditional numerical modeling and data-driven inference, offering a path toward real-time, uncertainty-aware simulation of complex porous systems. Key findings and contributions can be summarized as follows:

\begin{enumerate}[label=(\roman*)]
    \item Our modeling framework enables direct solution of the physical problem across a defined parameter space, effectively extending the PINN-UU concept \citep{panahi2025modeling} to spatially heterogeneous parameter fields. We demonstrate that a single neural network can learn to approximate heterogeneous Darcy flow solutions across a continuous range of governing parameters.

    \item We introduce and validate two distinct strategies for the differentiable parameterization of spatial heterogeneity of model parameters: (1) a direct functional approach for simple geometric features, and (2) a novel, data-driven approach using a pre-trained, coordinate-based autoencoder.

    \item A key technical innovation of the study is the direct integration of the autoencoder's differentiable decoder based on an Implicit Neural Representation (INR) into the PINN loss function. This enables on-the-fly computation of the conductivity field and its spatial derivatives via automatic differentiation (AD), which constitutes a critical step for enforcing the physics-based PDE residual.

    \item Through an extensive set of numerical analyses, we demonstrate that the learned solutions are robust and consistent with physical principles, such as local mass conservation, even as training is performed solely on the global PDE residuals. This highlights the ability of our framework to internalize the underlying physics.

    \item Computational efficiency of the trained model is documented to enable practical uncertainty quantification (UQ) tasks, such as those related to simulation of transport upon relying on a large-ensemble of particle tracking evaluations. 
\end{enumerate}

While our framework establishes a robust differentiable solver for parameterized Darcy flow, it also points toward several future directions. These include, for example, extending this methodology to transient flow and transport scenarios. Further investigations into more advanced generative models for encoding complex geological structures, for example moving beyond the current INR-decoder to consider variational or generative adversarial approaches, constitute an interesting avenue for future research. Finally, trained solution operator can serve as a robust prior that can be rapidly fine-tuned or conditioned on sparse, possibly noisy, experimental observations using (stochastic) inverse modeling techniques.

\section*{Declarations}
\label{section_declaration}

\textit{Funding.} The authors acknowledge financial support from the European Union’s Horizon 2020 research and innovation programme under the Marie Skłodowska-Curie grant agreement No 956384\\\\
\textit{Code availability.} Codes used in this paper will be available in the following github repository \url{https://github.com/MiladPnh/DarcyPINN-CausalTraining.git}\\\\
\textit{Author Contributions.}
M.P. and G.P came up with the research idea, designed the computation experiments, and M.P. implemented the code. G.P., A.G, and M.R. supervised the project. All participated in reviewing the manuscript.

\newpage
\bibliography{References}

\newpage
\appendix

\section*{Supporting information}
\setcounter{equation}{0}
\renewcommand{\theequation}{S\arabic{equation}}
\setcounter{figure}{0}
\renewcommand{\thefigure}{S\arabic{figure}}
\setcounter{table}{0}
\renewcommand{\thetable}{S\arabic{table}}
\section{PirateNet Architecture Details}
\label{app:piratenet}
\begin{figure*}[!ht]
\centering
\includegraphics[width=.9\textwidth]{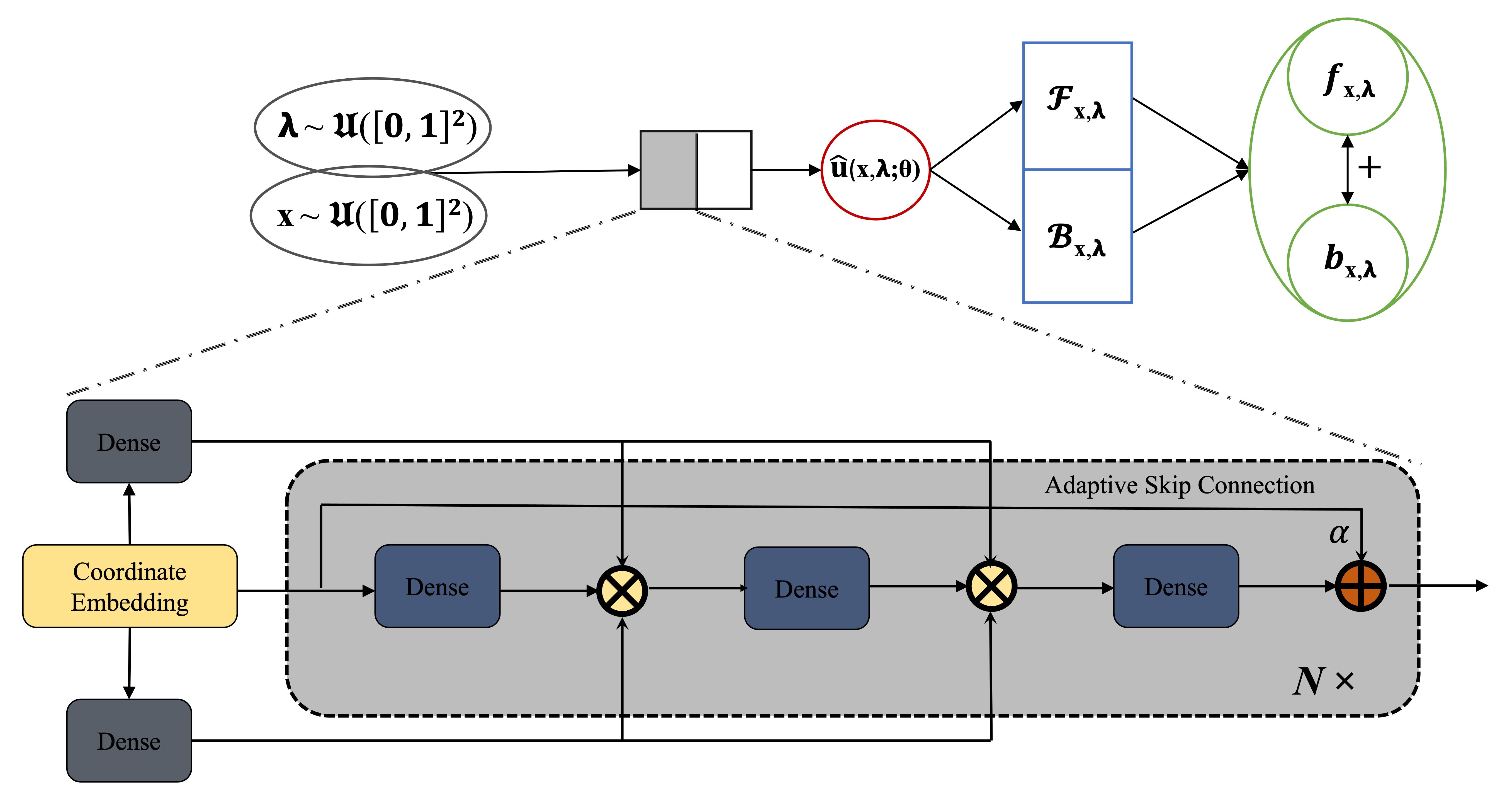}
\caption{\textbf{Schematic of the parameterized Physics-Informed Neural Network (PINN) architecture based on PirateNet for solving the steady-state Darcy flow problem.} 
The network takes spatial coordinates $\mathbf{x} \sim \mathcal{U}([0,1]^2)$ and parameter vectors $\boldsymbol{\lambda} \sim \mathcal{U}([0,1]^2)$ (representing, e.g., the center of a Gaussian conductivity feature) as input. 
The input undergoes coordinate embedding before being processed by two adaptive residual blocks (\textit{N}$\times$, where \textit{N}=2 in this study), detailed in the zoomed-in view. 
The output $u_{\boldsymbol{\Theta}_\lambda}$ is then used to evaluate the physics-informed loss, which includes residuals of the governing PDE operator $\mathcal{F}_{\mathbf{x},\boldsymbol{\lambda}}$ and boundary condition operator $\mathcal{B}_{\mathbf{x},\boldsymbol{\lambda}}$, compared against source terms $f_{\mathbf{x},\boldsymbol{\lambda}}$ and boundary values $b_{\mathbf{x},\boldsymbol{\lambda}}$, respectively. The adaptive skip connection within each PirateNet block is controlled by a trainable scalar $\alpha$.}
\label{fig:pinn_architecture_gaussian_case1} 
\end{figure*}

The core of our modeling framework is a parameterized Physics-Informed Neural Network (PINN) whose architecture, depicted in Figure~\ref{fig:pinn_architecture_gaussian_case1}, is inspired by the PirateNet design \cite{wang2024piratenets}. This design was chosen for its demonstrated ability to facilitate stable and efficient training of deep neural networks in the context of physics-informed learning. The architecture can be broken down into three main components: an input embedding layer, a series of adaptive residual blocks, and a final output layer.

The network is structured to accept both spatial coordinates $\mathbf{x}$ (sampled uniformly from the unit square) and a low-dimensional parameter latent vector $\boldsymbol{\lambda}$ as concatenated input. 
This combined input first passes through an initial dense layer for coordinate embedding, followed by processing through a series of adaptive residual blocks. As detailed in the magnified portion of Figure~\ref{fig:pinn_architecture_gaussian_case1}, each PirateNet block features internal dense layers, gating mechanisms based on the initial embedding, and a crucial adaptive skip connection modulated by a trainable scalar $\alpha$. 
This architecture facilitates the learning of a "solution bundle", where the network's output $u_{\boldsymbol{\Theta}_\lambda}$, is a function of both space and the latent space parameters defining the hydraulic conductivity spatial structure. 
The network is trained by minimizing a loss function derived from the residuals of the Darcy flow governing equations ($\mathcal{F}_{\mathbf{x},\boldsymbol{\lambda}}$) and boundary conditions ($\mathcal{B}_{\mathbf{x},\boldsymbol{\lambda}}$), ensuring physical consistency across the entire spatio-parameter domain without reliance on pre-computed simulation data.

\paragraph{Input Embedding with Random Fourier Features.}
To mitigate the well-known spectral bias of Multilayer Perceptrons (MLPs), the normalized input coordinates $(\mathbf{x}, \boldsymbol{\lambda})$ are mapped to a higher-dimensional feature space using a Random Fourier Feature (RFF) embedding. We define the combined input vector $\mathbf{z} = (\mathbf{x}, \boldsymbol{\lambda})$, where $\mathbf{z} \in [0,1]^{d_z}$ and $d_z = d_x + d_{\lambda}$. The embedding is given by:
\begin{equation}
    \Phi(\mathbf{z}) = \left[ \cos(2\pi \mathbf{B}\mathbf{z}), \sin(2\pi \mathbf{B}\mathbf{z}) \right]^T,
\end{equation}
where the entries of $\mathbf{B} \in \mathbb{R}^{m \times d_z}$ are sampled i.i.d. from an isotropic Gaussian distribution $\mathcal{N}(0 , 1)$ and fixed during training. This embedding enhances the network's ability to approximate solutions with rich spatial variability, overcoming the tendency of standard MLPs to over-smooth the output. The output of this layer, denoted $\Phi(\mathbf{z})$, is then passed to the main body of the network.

Following the embedding, we compute two global gating layers, $\mathbf{U}$ and $\mathbf{V}$, via two dense layers:
\begin{equation}
    \mathbf{U} = \sigma(\mathbf{W}_{u}\Phi(\mathbf{z}) + \mathbf{b}_{u}), \quad \mathbf{V} = \sigma(\mathbf{W}_{v}\Phi(\mathbf{z}) + \mathbf{b}_{v}),
\end{equation}
where $\sigma$ is the activation function. These two encoding maps act as global gates in each residual block of the architecture, that modulate the information flow based on the inputs $\mathbf{z}$ \citep{wang2021understanding, anagnostopoulos2023residualbasedattentionconnectioninformation}.

\paragraph{Adaptive Residual Blocks.}
The network comprises $L$ residual blocks. Let $\mathbf{z}^{(l)}$ denote the input to the $l$-th block (for $1 \le l \le L$) and $\mathbf{z}^{(0)} = (\mathbf{x}, \boldsymbol{\lambda})$ as initial input. The forward pass through the block matches the formulation in \citep{wang2024piratenets}, adapted here for our parameterized input $\mathbf{z}$. It involves three dense operations and two gating steps:
\begin{align}
    \mathbf{f}^{(l)} &= \sigma(\mathbf{W}_{1}^{(l)} \mathbf{z}^{(l)} + \mathbf{b}_{1}^{(l)}) \label{eq:pirate1} \\
    \mathbf{m}_{1}^{(l)} &= \mathbf{f}^{(l)} \odot \mathbf{U} + (1 - \mathbf{f}^{(l)}) \odot \mathbf{V} \label{eq:pirate2} \\
    \mathbf{g}^{(l)} &= \sigma(\mathbf{W}_{2}^{(l)} \mathbf{m}_{1}^{(l)} + \mathbf{b}_{2}^{(l)}) \label{eq:pirate3} \\
    \mathbf{m}_{2}^{(l)} &= \mathbf{g}^{(l)} \odot \mathbf{U} + (1 - \mathbf{g}^{(l)}) \odot \mathbf{V} \label{eq:pirate4} \\
    \mathbf{n}^{(l)} &= \sigma(\mathbf{W}_{3}^{(l)} \mathbf{m}_{2}^{(l)} + \mathbf{b}_{3}^{(l)}) \label{eq:pirate5} \\
    \mathbf{z}^{(l+1)} &= \alpha^{(l)} \mathbf{n}^{(l)} + (1 - \alpha^{(l)}) \mathbf{z}^{(l)} \label{eq:pirate6}
\end{align}
Here, $\odot$ denotes element-wise multiplication, $\sigma$ is a non-linear activation function (e.g., Tanh), and the terms $\mathbf{W}$ and $\mathbf{b}$ represent trainable weight matrices initialized by the Glorot scheme \citep{glorot2010understanding} and bias vectors, respectively. We introduce the symbol $\mathbf{n}^{(l)}$ (Eq. \ref{eq:pirate5}) to represent the block's non-linear mapping, distinguishing it from the hydraulic head $h$.

\paragraph{Final Output Layer and Initialization Properties.}
The output of the final residual block, $\mathbf{z}^{(L+1)}$, is projected to the physical solution variable (the hydraulic head in this work) via a final linear dense layer:
\begin{equation}
    u_{\boldsymbol{\Theta}_\lambda}(\mathbf{x}, \boldsymbol{\lambda}) = \mathbf{W}_{(L+1)} \mathbf{z}^{(L+1)} + b_{out},
\end{equation}
where $\mathbf{W}_{(L+1)}$ and $b_{(L+1)}$ are the final learnable weight matrix and bias.

The key innovation lies in Eq. \eqref{eq:pirate6}, where $\alpha^{(l)} \in \mathbb{R}$ is a trainable scalar parameter. A defining characteristic of the PirateNet architecture is its behavior at the start of training. Since the skip-connection parameters $\alpha^{(l)}$ are initialized to zero, every residual block initially functions as an identity map (i.e., $\mathbf{z}^{(l+1)} = \mathbf{z}^{(l)}$). By induction, the input to the final layer at initialization is identical to the output of the first embedding layer, $\mathbf{z}^{(L+1)} = \Phi(\mathbf{z})$. Consequently, at step zero, the entire deep network collapses mathematically into a linear combination of the first layer embeddings at initialization:
\begin{equation}
    u_{\boldsymbol{\Theta}_\lambda}^{\text{init}}(\mathbf{z}) = \mathbf{W}_{(L+1)} \Phi(\mathbf{z}) + b_{(L+1)}.
\end{equation}
This property is significant for two reasons. First, it circumvents the "initialization pathologies" often observed in deep PINNs, where random weight initialization in deep non-linear networks can lead to vanishing gradients or stiff optimization landscapes. By starting as a shallow linear model, the network ensures stable gradient flow. As training progresses and $\alpha^{(l)}$ evolves, the network progressively "unlocks" its depth and non-linearity, increasing its expressivity only as required to minimize the PDE residuals.

Second, this formulation allows the network to be viewed as a linear expansion of basis functions at initialization. This structure enables the optional integration of prior knowledge or available data (e.g., boundary conditions or sparse measurements $\mathbf{Y}$) directly into the initialization phase. One can solve for the optimal initial weights $\mathbf{W}_{out}$ by minimizing a standard least-squares objective:
\begin{equation}
    \min_{\mathbf{W}_{out}} \| \mathbf{W}_{out} \Phi(\mathbf{z}) - \mathbf{Y} \|_2^2.
\end{equation}
This capability provides a mathematically robust "initial guess" for the solver, akin to spectral methods, before the physics-informed optimization begins fine-tuning the non-linear parameters.

\section{Autoencoder Data Generation and Architecture}
\label{app:ae}
This Supporting Information provides supplementary details regarding the generation of training data and the specific architecture of the autoencoder (AE) used in Scenario 2 (Section \ref{sec:autoencoder_param}). The AE is employed to derive a low-dimensional parameterization $\boldsymbol{\lambda}$ for imposing the spatial distribution fo hydraulic conductivity $K(\mathbf{x})$, which is then used as input to the main physics-informed neural network (PINN) solver.

\subsection{Gaussian Random Field (GRF) Sample Generation}
\label{app:grf_generation}

The training dataset for the autoencoder consists of $M$ realizations of hydraulic conductivity, generated as two-dimensional Random Fields. We utilize a spectral method based on the Fast Fourier Transform (FFT) for efficient generation. This method ensures the generated fields possess specified statistical properties, primarily governed by their power spectral density (PSD).

The generation process for each sample on an $N \times N$ grid (with $N=32$ in our setup) involves the following steps:
\begin{enumerate}[label=(\roman*)]
    \item \textbf{Frequency Grid:} Define the discrete spatial frequencies $(k_x, k_y)$ corresponding to the $N \times N$ grid using \texttt{FFT}.
    \item \textbf{Power Spectral Density (PSD):} Define the PSD, $S(k_x, k_y)$, corresponding to a desired covariance model. For a Gaussian covariance function with correlation length $L=0.15$ , the PSD is given by:
        \begin{equation}
            S(k_x, k_y) = C \exp\left(-2 (\pi L)^2 (k_x^2 + k_y^2)\right)
            \label{eq:app_psd}
        \end{equation}
        where $C$ is a normalization constant.
    \item \textbf{Random Fourier Coefficients:} Generate a realization of complex Gaussian white noise $\eta(k_x, k_y)$ in the frequency domain, where the real and imaginary parts are drawn independently from $\mathcal{N}(0, 1)$.
    \item \textbf{Field in Fourier Domain:} Construct the Fourier representation of the GRF, $K_{FFT}(k_x, k_y)$, by scaling the noise with the square root of the PSD:
        \begin{equation}
            K_{FFT}(k_x, k_y) = \eta(k_x, k_y) \sqrt{S(k_x, k_y) / 2}
            \label{eq:app_kfft}
        \end{equation}
        The division by 2 accounts for allocating power to both positive and negative frequencies for a real-valued field. The zero-frequency component $K_{FFT}(0, 0)$ is set to zero to enforce a zero mean for the generated field.
    \item \textbf{Inverse FFT:} Obtain the real-space GRF sample $K(\mathbf{x})$ by applying the inverse Fast Fourier Transform (IFFT) and taking the real part:
        \begin{equation}
            K(\mathbf{x}) = \text{Re}\left[ \text{IFFT}_{2D} \{ K_{FFT}(k_x, k_y) \} \right]
            \label{eq:app_grf_ifft}
        \end{equation}
\end{enumerate}
This procedure is repeated $M$ times to generate the dataset $\{K^{(m)}\}_{m=1}^M$.

\subsection{Data Preprocessing for Autoencoder Training}
\label{app:data_preprocessing}

Before being fed into the autoencoder, the raw GRF samples undergo several preprocessing steps as implemented in the code:
\begin{enumerate}[label=\roman*.]
    \item \textbf{Standardization:} Global mean $\mu$ and standard deviation $\sigma$ are computed across all pixels of all scaled samples. The fields are then standardized: $K_{std}^{(m)} = (K_{scaled}^{(m)} - \mu) / (\sigma + \epsilon)$, where $\epsilon$ is a small constant for numerical stability.
    \item \textbf{Normalization to [0, 1]:} Global minimum $K_{min,std}$ and maximum $K_{max,std}$ are found across all standardized samples. The fields are then normalized to the range $[0, 1]$:
        \begin{equation}
            K_{norm}^{(m)} = \frac{K_{std}^{(m)} - K_{min,std}}{K_{max,std} - K_{min,std} + \epsilon}
            \label{eq:app_norm01}
        \end{equation}
    This final normalization step produces the input data $K_{norm}^{(m)}$ for the autoencoder. A channel dimension is added, resulting in input tensors of shape $[M, N, N, 1]$.
\end{enumerate}

\subsection{Autoencoder Architecture}
\label{app:autoencoder}
This Supporting Information provides further technical details on the autoencoder architecture employed in Scenario 2 (Section \ref{sec:autoencoder_param}) for generating parameterized hydraulic conductivity tensors $K(\mathbf{x}; \boldsymbol{\lambda})$ within the physics-informed neural network (PINN) framework. The autoencoder serves to learn a low-dimensional latent representation $\boldsymbol{\lambda}$ from a dataset of high-dimensional Gaussian Random Field (GRF) samples, enabling efficient parameterization of complex geological structures.

The autoencoder consists of an encoder $E_{\phi_1}$ and an Implicit Neural Representation (INR) decoder (coordinate-based decoder) $D_{\phi_2}$.

\subsubsection{Encoder Architecture}
The encoder network $E_{\phi_1}$ maps an input conductivity field realization $K$, typically discretized on an $N \times N$ grid (input shape $[N, N, 1]$), to a low-dimensional latent vector $\boldsymbol{\lambda} = E_{\phi_1}(K)$. The architecture adheres to a standard Convolutional Neural Network (CNN) design typical of autoencoder frameworks. It comprises a series of convolutional layers using $3 \times 3$ kernels with a stride of 2. This stride configuration performs spatial downsampling, effectively halving the grid resolution at each step while simultaneously increasing the number of feature channels. Normalization layers and ReLU activation functions are applied after each convolution to stabilize training and introduce non-linearity. 

Upon passing through the convolutional blocks, the resulting feature map is flattened (i.e., reshaped into a one-dimensional vector) and processed by a sequence of fully connected dense layers with ReLU activations. The final layer projects this vector down to the target latent dimension $d_\lambda$ via a linear activation. In this study, we set $d_\lambda=2$. This choice represents a deliberate trade-off: while higher dimensions would increase the representational power of the autoencoder, we prioritized a compact parameterization to ensure computational tractability for the downstream PINN solver. We selected the minimum dimension sufficient to capture the essential heterogeneity, thereby  limiting the dimensionality of the PINN's input space while also facilitating direct visualization. The weights and biases of these layers constitute the parameter set $\phi_1$.

\subsubsection{INR (coordinate-based) Decoder Architecture}
\label{app:decoder}

The INR decoder network $D_{\phi_2}$ adopts a resolution-invariant approach, differing from typical deconvolutional architectures. Instead of reconstructing a fixed-size grid, it functions as an implicit neural representation \cite{sitzmann2020implicit}. It takes a specific latent vector $\boldsymbol{\lambda} \in \mathbb{R}^{d_\lambda}$ and a set of arbitrary spatial query coordinates $\mathbf{X}_{query} = \{\mathbf{x}_j = (x_j, y_j)\}_{j=1}^{N_q}$ as input, and outputs the corresponding predicted conductivity values $\{ \hat{K}(\mathbf{x}_j; \boldsymbol{\lambda}) \}_{j=1}^{N_q}$ at those locations. This design is particularly advantageous as it allows the trained decoder to generate conductivity fields at any desired spatial resolution during the subsequent PINN training phase, independent of the resolution used for autoencoder training. The architecture combines positional encoding of coordinates with a Multi-Layer Perceptron (MLP).

\paragraph{Positional Encoding of Coordinates.}

To effectively incorporate spatial information into the MLP, the continuous input coordinates $\mathbf{x} = (x, y)$ (normalized to $[0, 1]$) are first mapped to a higher-dimensional feature space using a positional encoding (PE) function $\gamma(\cdot)$, similar to techniques used in NeRF \cite{tancik2020fourier} and Transformers \cite{vaswani2017attention}. This encoding helps the MLP capture high-frequency details in the spatial domain. Following the implementation, we define a set of $N_{PE}$ frequency multipliers (bands) $B = \{b_k\}_{k=1}^{N_{PE}}$. For a single coordinate component $x$, the PE is calculated as:
\begin{equation}
    \gamma(x) = \left[ \cos(2 \pi b_1 x), \sin(2 \pi b_1 x), \dots, \cos(2 \pi b_{N_{PE}} x), \sin(2 \pi b_{N_{PE}} x) \right]^T \in \mathbb{R}^{2 N_{PE}}
    \label{eq:app_pe}
\end{equation}
The full positional encoding for the coordinate vector $\mathbf{x}=(x,y)$ is obtained by concatenating the encodings for each component:
\begin{equation}
    \gamma(\mathbf{x}) = \begin{bmatrix} \gamma(x) \\ \gamma(y) \end{bmatrix} \in \mathbb{R}^{4 N_{PE}}
    \label{eq:app_pe_xy}
\end{equation}
This results in a fixed-size, high-dimensional representation of the spatial location, irrespective of the number of query points.

\paragraph{MLP for Conductivity Prediction.}

For each query coordinate $\mathbf{x}_j$ within a batch associated with a latent vector $\boldsymbol{\lambda}$, the INR decoder first computes its positional encoding $\gamma(\mathbf{x}_j)$. This encoded coordinate vector is then concatenated with the latent vector $\boldsymbol{\lambda}$. This combined feature vector serves as the input to a deep MLP:
\begin{equation}
    \mathbf{z}_j = \text{concat}(\boldsymbol{\lambda}, \gamma(\mathbf{x}_j)) \in \mathbb{R}^{d_\lambda + 4 N_{PE}}
    \label{eq:app_mlp_input}
\end{equation}
The MLP, comprising the bulk of the decoder parameters $\phi_2$, consists of a sequence of dense layers with ReLU activation functions, transforming the input $\mathbf{z}_j$ through progressively learned feature representations $\mathbf{o}_j^{(l)}$:
\begin{equation}
    \mathbf{o}_j^{(l+1)} = \text{ReLU}(\mathbf{W}^{(l)} \mathbf{o}_j^{(l)} + \mathbf{b}^{(l)}), \quad \text{with } \mathbf{o}_j^{(0)} = \mathbf{z}_j
\end{equation}
The specific layer widths used in our implementation ($[8, 32, 128, 256, 256, 128, 32, 8]$ neurons per layer) define the capacity of the network to model the mapping from the latent space and coordinates to the conductivity value.

\paragraph{Denormalization.}

A final dense layer with a single output neuron and a suitable activation function $\sigma_{out}$ (`hard\_sigmoid` as implemented, which clips the output to $[0, 1]$) produces the normalized predicted conductivity value:
\begin{equation}
     \hat{K}_{norm}(\mathbf{x}_j; \boldsymbol{\lambda}) = \sigma_{out}(\mathbf{W}_{out} \mathbf{o}_j^{(L)} + b_{out}) \in [0, 1]
     \label{eq:app_k_norm}
\end{equation}
where $L$ is the number of hidden layers in the MLP.

Finally, the normalized output $\hat{K}_{norm}$ is mapped back to the physical domain. We note that while the preprocessing (Section \ref{app:data_preprocessing}) consists of sequential standardization and normalization steps, both are linear operations. Consequently, their inverse is applied here as a single composite affine transformation that maps the network output directly from $[0, 1]$ to the physical range of the hydraulic conductivity $[K_{min}, K_{max}]$:
\begin{equation}
     \hat{K}(\mathbf{x}_j; \boldsymbol{\lambda}) = (K_{max} - K_{min}) \cdot \hat{K}_{norm}(\mathbf{x}_j; \boldsymbol{\lambda}) + K_{min}
     \label{eq:app_k_denorm}
\end{equation}
In our specific experimental setup, where the bounds are defined by $K_{min}=e^{-0.8}$ and $K_{max}=e^{0.8}$, this yields:
\begin{equation}
     \hat{K}(\mathbf{x}_j; \boldsymbol{\lambda}) = (e^{0.8} - e^{-0.8}) \cdot \hat{K}_{norm}(\mathbf{x}_j; \boldsymbol{\lambda}) + e^{-0.8}.
\end{equation}
This final value $\hat{K}(\mathbf{x}_j; \boldsymbol{\lambda})$ is the hydraulic conductivity used by the PINN solver when evaluating the physics-informed loss for parameter vector $\boldsymbol{\lambda}$ at location $\mathbf{x}_j$.

\subsection{Autoencoder Training Objective}
\label{app:ae_loss}

The autoencoder parameters $\phi = (\phi_1, \phi_2)$ are optimized jointly by minimizing a composite loss function $\mathcal{L}_{AE}$. This function is designed to ensure accurate reconstruction while preserving key statistical and structural properties of the input fields. The total loss is defined as a weighted sum of five components:
\begin{equation}
    \mathcal{L}_{AE} = w_{recon}\mathcal{L}_{recon} + w_{mean}\mathcal{L}_{mean} + w_{var}\mathcal{L}_{var} + w_{freq}\mathcal{L}_{freq} + w_{smooth}\mathcal{L}_{smooth}
    \label{eq:app_total_ae_loss}
\end{equation}
where the weights are set to $(w_{recon}, w_{mean}, w_{var}, w_{freq}, w_{smooth}) = (50, 50, 50, 10, 50)$ in our implementation. The individual components are defined as follows:

\paragraph{Weighted Reconstruction Loss ($\mathcal{L}_{recon}$).}
We employ a modified Mean Squared Error (MSE) that emphasizes accuracy near the extrema (minimum and maximum values) of the conductivity field. We first define a spatial weight map $w^{(m)}(\mathbf{x})$ inversely proportional to the distance from the global bounds:
\begin{equation}
    w^{(m)}(\mathbf{x}) \propto \frac{1}{\epsilon + \min\left(|K_{norm}^{(m)}(\mathbf{x})-K_{min}^{(m)}|, |K_{norm}^{(m)}(\mathbf{x})-K_{max}^{(m)}|\right)},
\end{equation}
which is normalized such that its spatial mean is 1. The loss is then the weighted MSE:
\begin{equation}
    \mathcal{L}_{recon} = \mathbb{E}\left[w^{(m)}(\mathbf{x}) \left(\hat{K}_{norm}^{(m)}(\mathbf{x}) - K_{norm}^{(m)}(\mathbf{x})\right)^2\right].
\end{equation}

\paragraph{Statistical Moment Losses ($\mathcal{L}_{mean}$ and $\mathcal{L}_{var}$).}
To ensure the generative model captures the global statistics of the field, we penalize deviations in the spatial mean ($\mu$) and variance ($\sigma^2$):
\begin{align}
    \mathcal{L}_{mean} &= \mathbb{E}\left[(\mu_{K^{(m)}} - \mu_{\hat{K}^{(m)}})^2\right], \\
    \mathcal{L}_{var} &= \mathbb{E}\left[(\sigma^2_{K^{(m)}} - \sigma^2_{\hat{K}^{(m)}})^2\right].
\end{align}

\paragraph{Frequency Domain Loss ($\mathcal{L}_{freq}$).}
This term ensures the spectral properties of the field are preserved by minimizing the difference in the magnitude of the 2D Fast Fourier Transform (FFT) coefficients:
\begin{equation}
    \mathcal{L}_{freq} = \mathbb{E}\left[|\text{FFT}(\hat{K}_{norm}^{(m)}) - \text{FFT}(K_{norm}^{(m)})|\right].
\end{equation}

\paragraph{Smoothness Loss ($\mathcal{L}_{smooth}$).}
To encourage physical consistency and spatial smoothness, we penalize the magnitude of the first- and second-order spatial gradients (Laplacian smoothness) of the reconstructed field:
\begin{equation}
    \mathcal{L}_{smooth} = \|\nabla \hat{K}_{norm}^{(m)}\|^2 + \|\nabla^2 \hat{K}_{norm}^{(m)}\|^2.
\end{equation}
The optimization is performed using the Adam optimizer \cite{kingma2014adam} with an exponential decay learning rate schedule.

\subsubsection{PINN Solver with Explicit Conditioning Parameterization Scheme}
\label{sec:pinn_latent}

Once the Autoencoder is trained, its parameters are stored, particularly those related to the INR decoder $D_{\phi_2}$. These parameters are needed to train the PirateNet-based PINN solver $u_{\boldsymbol{\Theta}_\lambda}$.
In Scenario 2, the parameter vector for the PINN is the latent vector $\boldsymbol{\lambda} \in \Lambda \subset \mathbb{R}^{d_\lambda}$. We typically sample $\boldsymbol{\lambda}$ from the distribution induced by the encoder on the training data, $p(\boldsymbol{\lambda}) \approx p(E_{\phi_1}(K))$, or a fitted approximation (e.g., a multivariate Gaussian distribution fitted to the encoded vectors $\{\boldsymbol{\lambda}^{(m)}\}_{m=1}^M$, as visualized in Figure \ref{fig:architecture_autoencoder_case2_method}).

The input to the PirateNet solver $u_{\boldsymbol{\Theta}_\lambda}$ is the concatenated vector $(x, y, \boldsymbol{\lambda})$. Crucially, the hydraulic conductivity field $K(\mathbf{x}; \boldsymbol{\lambda})$ required within the PINN's loss function (Eq. \ref{eq:total_loss_detailed}) is now defined by the pre-trained decoder:
\begin{equation}
    K(\mathbf{x}; \boldsymbol{\lambda}) = \tilde{D}_{\phi_2}(\boldsymbol{\lambda}, \mathbf{x})
    \label{eq:k_decoder}
\end{equation}
where $\tilde{D}_{\phi_2}$ represents the decoder function, including the final denormalization step to scale the output to the physical conductivity range (e.g., $[e^{-.8}, e^{.8}]$).

During the PINN training (Section \ref{sec:training_strategy}), for a sampled point $(\mathbf{x}_i, \boldsymbol{\lambda}_i)$:
\begin{enumerate}[label=\roman*]
    \item $K(\mathbf{x}_i; \boldsymbol{\lambda}_i)$ is computed by passing $\boldsymbol{\lambda}_i$ and $\mathbf{x}_i$ through the decoder $D_{\phi_2}$ (Eq.~\ref{eq:k_decoder}).
    \item The spatial derivatives $\nabla_{\mathbf{x}} K(\mathbf{x}_i; \boldsymbol{\lambda}_i)$ needed for the PDE residual are obtained by applying AD through the computational graph of the decoder $D_{\phi_2}$ with respect to the spatial inputs $\mathbf{x}_i$.
\end{enumerate}
This approach allows the PINN solver $u_{\boldsymbol{\Theta}_\lambda}$ to learn the mapping from the low-dimensional latent representation $\boldsymbol{\lambda}$ and spatial coordinates $\mathbf{x}$ to the corresponding head and velocity fields. The use of the INR decoder $D_{\phi_2}$ renders the definition of $K$ continuous and independent of the discretization mesh. This implies the PINN can be trained on arbitrary collocation points not bound to the original training grid, while the spectral content of the reconstructed field remains effectively bounded by the resolution of the data used to train the autoencoder.

\section{Physics-Informed Neural Network (PINN) Framework}
\label{app:pinn_framework}
This section provides technical details on the parameterized PINN framework used in this work. We formalize the differentiable solver ansatz, the composition of the physics-informed loss function, and the specific PirateNet-inspired network architecture that enables stable training of deep networks.

\subsection{The Differentiable Solver Ansatz}
Our framework approximates the solution operator by learning the scalar hydraulic head field, $\hat{h}$, as a primary output. A single, unified neural network, $u_{\boldsymbol{\Theta}_\lambda}$, parameterized by trainable weights and biases $\boldsymbol{\theta}$, is defined to map the spatial coordinates and input parameters $(\mathbf{x}, \boldsymbol{\lambda})$ to the predicted head
\begin{equation}
    \hat{h}(\mathbf{x}; \boldsymbol{\lambda}) = u_{\boldsymbol{\Theta}_\lambda}(\mathbf{x}, \boldsymbol{\lambda}).
    \label{eq:app_nn_ansatz}
\end{equation}
The corresponding velocity field, $\hat{\mathbf{v}}$, is not a direct output of the network. Instead, it is \textit{derived} from the head output by applying automatic differentiation (AD) to enforce Darcy's law (Eq.~\ref{eq:strong_darcy_method}):
\begin{equation}
    \hat{\mathbf{v}}(\mathbf{x}; \boldsymbol{\lambda}) := -K(\mathbf{x}; \boldsymbol{\lambda}) \nabla_{\mathbf{x}} \hat{h}(\mathbf{x}; \boldsymbol{\lambda}).
    \label{eq:app_velocity_derivation}
\end{equation}
This formulation ensures that the learned velocity field is, by construction, divergence-free with respect to the learned head field and the conductivity field, a critical aspect of the differentiable physics paradigm.

\subsection{Physics-Informed Loss Function}
The network's parameters $\boldsymbol{\theta}$ are optimized by minimizing a composite loss function, $\mathcal{J}(\boldsymbol{\theta})$, constructed from the residuals of the governing physical laws. Given that the velocity $\hat{\mathbf{v}}$ is derived from $\hat{h}$, the domain loss comprises the continuity equation residual and the boundary condition residuals.

The continuity residual, $r_{\text{cont}}$, measures the extent to which the derived velocity field violates mass conservation:
\begin{equation}
    r_{\text{cont}}(\mathbf{x}, \boldsymbol{\lambda}; \boldsymbol{\theta}) := \nabla \cdot \hat{\mathbf{v}}(\mathbf{x}; \boldsymbol{\lambda}) + f(\mathbf{x}).
\end{equation}
The boundary residual, $r_{\text{bc}}$, measures the mismatch on the domain boundary $\partial\Omega$. For a Dirichlet boundary condition of the form $h(\mathbf{x}) = g_D(\mathbf{x})$, the residual is $r_{\text{bc}} = \hat{h}(\mathbf{x}; \boldsymbol{\lambda}) - g_D(\mathbf{x})$. For a Neumann boundary condition involving a prescribed flux $g_N$, the residual is $r_{\text{bc}} = \hat{\mathbf{v}}(\mathbf{x}; \boldsymbol{\lambda}) \cdot \mathbf{n} - g_N$, where $\mathbf{n}$ is the outward normal vector. All spatial derivatives applied to the head field required to compute these residuals are evaluated via AD.

The total loss is a weighted sum of the mean squared norms of these residuals, approximated by sampling $N_r$ collocation points $\{(\mathbf{x}_i, \boldsymbol{\lambda}_i)\}$ from the domain $\Omega \times \Lambda$ and $N_b$ from the boundary $\partial\Omega \times \Lambda$:
\begin{equation}
    \mathcal{J}(\boldsymbol{\theta}) = w_{\text{cont}} \mathcal{J}_{\text{cont}} + w_{\text{bc}} \mathcal{J}_{\text{bc}},
    \label{eq:app_total_loss}
\end{equation}
where
\begin{align}
    \mathcal{J}_{\text{cont}} &= \frac{1}{N_r} \sum_{i=1}^{N_r} (r_{\text{cont}}(\mathbf{x}_i, \boldsymbol{\lambda}_i; \boldsymbol{\theta}))^2, \\
    \mathcal{J}_{\text{bc}} &= \frac{1}{N_b} \sum_{j=1}^{N_b} (r_{\text{bc}}(\mathbf{x}_j, \boldsymbol{\lambda}_j; \boldsymbol{\theta}))^2.
\end{align}
The loss weights, $w_{\text{cont}}$, $w_{\text{bc}}$ could be defined manually, or adaptively optimized.

\subsection{AutoGrad vs FDM}
\label{AD}

\begin{figure*}[!ht]
\centering
\includegraphics[width=1.\textwidth]{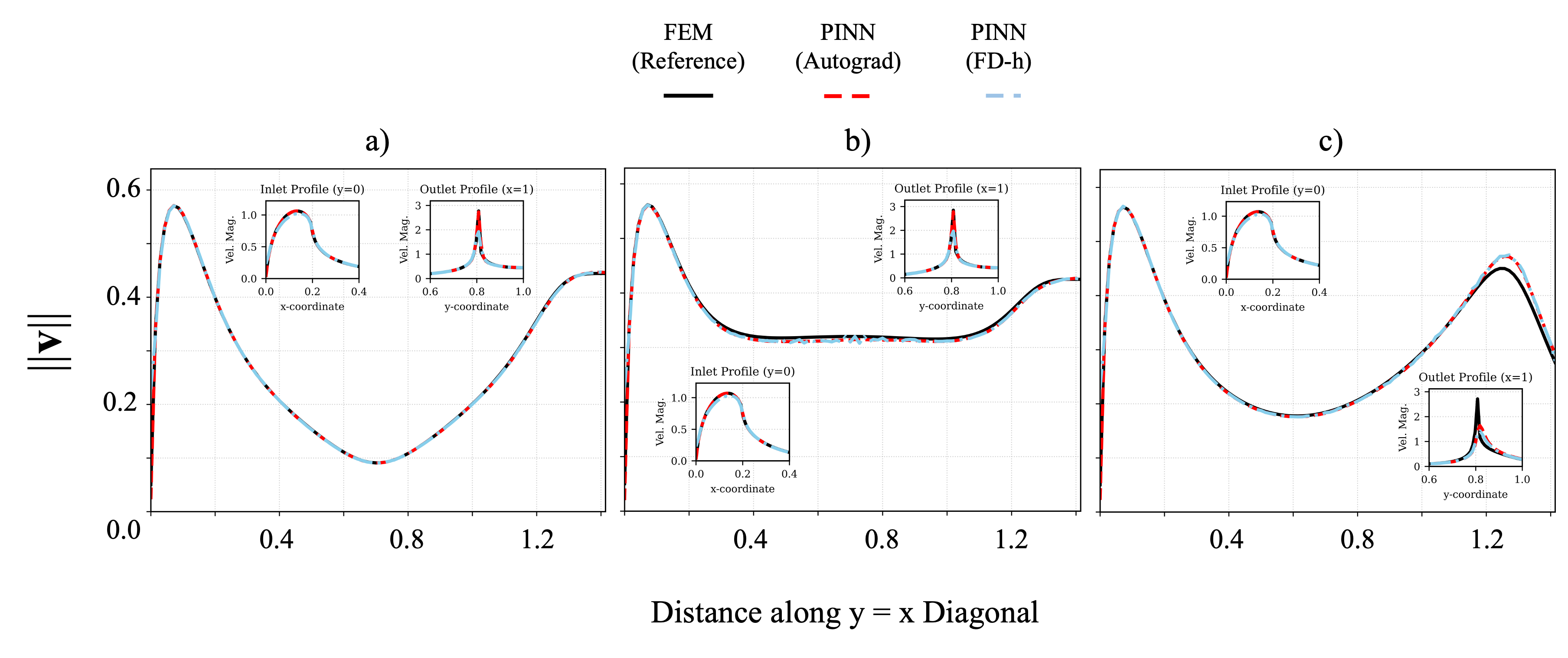}
\caption{\textbf{Comparison of velocity profiles obtained from FEM (reference, black solid line), PINN with velocity derived via Automatic Differentiation (Autograd, red dashed line), and PINN with velocity derived via Finite Differences from head output (FD-h, light blue dash-dotted line).} 
Profiles are shown along the main diagonal ($y=x$) of the domain for three different $\boldsymbol{\lambda}$ realizations corresponding to Gaussian bump centers at: (a) $\boldsymbol{\lambda}=(0.2, 0.8)$, (b) $\boldsymbol{\lambda}=(0.5, 0.5)$, and (c) $\boldsymbol{\lambda}=(0.8, 0.2)$. 
Insets show detailed velocity magnitude profiles at the inlet ($y=0$ or $x=0$) and outlet ($x=1$ or $y=1$) boundaries, corresponding to the flow direction for each case.}
\label{fig:velocity_profiles_comparison_gaussian_case1} 
\end{figure*}

\paragraph{Fidelity of Differentiable Velocity Fields.}
A key aspect of our framework is that the velocity field is not a direct network output but is derived from the head using automatic differentiation (AD). To investigate the fidelity of this derived field, we compare the velocity profiles obtained via AD against both the FEM reference and a velocity field calculated using numerical finite differences (FD) on the PINN's head output. As shown in Figure~\ref{fig:velocity_profiles_comparison_gaussian_case1}, the Autograd-derived velocity profiles exhibit excellent agreement with the FEM reference across all tested realizations. In contrast, the FD-derived velocities show more noticeable deviations, particularly in regions of sharp velocity gradients. The insets, detailing inlet and outlet profiles, further highlight the superior accuracy and smoothness of the Autograd approach, reinforcing the benefit of leveraging the network's differentiability to compute physically consistent velocity fields, a crucial aspect for accurately capturing flux and transport phenomena.

\section{Background on PINN Paradigms and Training Challenges}
\label{app:pinn_background}

\subsection{PINNs vs. PINOs: A Trade-off in Fidelity, Generality, and Data.}
The emergence of powerful neural operator frameworks leads to a key methodological crossroads in SciML: whether to learn the solution operator from data or to refine the solution-specific strategy of PINNs. An analysis by \citet{noguer2025mathematics} frames this as a trade-off between the \textit{solution-specific accuracy} of PINNs and the \textit{operator-level generalization} of Physics-Informed Neural Operators (PINOs) \citep{li2024physics}. PINNs excel at yielding high-fidelity solutions for specific problem instances by embedding the governing PDE directly into the loss function. In contrast, PINOs are designed for efficiency in parametric studies. They achieve this by training a neural operator on a combination of available input-output simulation data and a physics-based loss term computed on collocation points. The physics-based loss acts as a regularizer, enabling the PINO to learn from a sparser set of simulation data than a purely data-driven operator would require. However, this approach still prioritizes operator-level generalization and may sacrifice the pinpoint accuracy that a solution-specific PINN can achieve for any single instance.

This paper challenges the notion that one must choose between these paradigms. We explore how to advance the PINN framework itself to address two of its primary challenges. The first is the issue of parameter scalability: a standard PINN is architecturally a solver for a single problem instance and requires a complete, independent re-training for each new parameter value, making it inefficient for parametric studies. The second is the presence of known training pathologies, such as spectral bias and vanishing gradients, which can hinder convergence. By addressing these challenges, we aim to develop a framework that retains the high physical fidelity of a direct solver while achieving the parametric flexibility of a learned operator. This positions our work as a principled, physics-grounded alternative to purely data-driven regularization.

\subsection{The Pathologies of Training Deep Physics-Informed Models.}
The extension of PINNs to parameterized systems is complicated by two primary challenges. The first is an issue of \textbf{parameter scalability}: a standard PINN is architecturally a solver for a single problem instance and requires a complete, independent re-training for each new parameter value, making it inefficient for parametric studies.

Extension of PINNs to parameterized systems is further complicated by known training pathologies that are particularly acute in deep and complex network architectures \citep{glorot2010understanding}. A fundamental limitation is \textit{spectral bias}, wherein standard neural networks exhibit a preference for learning low-frequency components of a function, while documenting difficulties to represent the high-frequency features often present in solutions obtained as outputs of PDEs embedding governing physical laws (that might contain fine-scale spatial or temporal variations arising from sharp gradients and/or localized phenomena) \citep{Wang_2022, rahaman2019spectral, wang2021eigenvector}. Moreover, the optimization landscape of PINNs is notoriously difficult \citep{wang2022and}. The total loss function comprises different terms (e.g., PDE residuals, boundary conditions, and/or initial conditions) which often have vastly different magnitudes and associated gradient dynamics. This leads to unbalanced back-propagated gradients, hindering convergence and requiring sophisticated adaptive weighting schemes to manage these (often competing) objectives \citep{wang2021understanding, mcclenny2023self}. These challenges have inspired research into more robust network architectures \citep{wang2022improved}, such as those incorporating residual connections \citep{he2016deep, anagnostopoulos2023residual}. The PirateNet architecture, for instance, is specifically designed to address these initialization and training instability issues upon leveraging adaptive residual blocks that allow the network to dynamically adjust its effective depth during training \citep{wang2024piratenets}. In addition, advanced training strategies such as curriculum learning \citep{krishnapriyan2021characterizing, penwarden2023metalearning} and meta-learning approaches \citep{qin2022metapde, CHEN2022110996, Liu_2022, pmlrv70finn17a, antoniou2019trainmaml, nichol2018reptile}, provide promising avenues to advance the parametrized PDE solution paradigm.

\subsection{Theoretical Foundations.}
The theoretical foundation underpinning our approach can be viewed through the lens of the Implicit Function Theorem (IFT) and the principles of differentiable programming \citep{baydin2018automatic}. In this context, one can note that for any specific parameter instance $\boldsymbol{\lambda}_i$, an idealized PINN would find optimal weights $\boldsymbol{\theta}_i^*$ that nullify the gradient of the component of the loss associated with physics constraints, i.e., $\nabla_{\boldsymbol{\theta}} \mathcal{L}(\boldsymbol{\theta}_i^*; \boldsymbol{\lambda}_i) = 0$. The IFT provides the mathematical foundation to assess how this optimal solution (denoted as $u_{\boldsymbol{\theta}_i^*}$) co-varies with changes in $\boldsymbol{\lambda}_i$. While our framework does not compute the IFT's derivatives by differentiating through the underlying optimization process (a strategy employed in contexts such as quadratic programs and bilevel optimization \citep{amos2017optnet, fletcher2000practical}), it is instead tasked to learning a single global parameterization ($\boldsymbol{\theta}_{\text{global}}$) that approximates the solution manifold defined implicitly by the IFT.

\section{FEM as baseline}
\label{sec:appendix_fem}
Solutions obtained with the Finite Element Method (FEM) are considerd as a baseline reference to validate our PINN results. Our FE approach is based on a standard implementation, here recalled for completeness.
The FEM is based on the weak (or variational) formulation of the governing PDE system. By combining the continuity equation (Eq.~\ref{eq:strong_cont_method}) and Darcy's law (Eq.~\ref{eq:strong_darcy_method}), we obtain the second-order elliptic PDE for hydraulic head, $h$:
\begin{equation}
    -\nabla \cdot (K(\mathbf{x}; \boldsymbol{\lambda}) \nabla h(\mathbf{x}; \boldsymbol{\lambda})) = f(\mathbf{x}), \quad \forall \mathbf{x} \in \Omega.
\end{equation}
The weak form is derived by multiplying this equation by a suitable test function $\psi$ and integrating over the domain $\Omega$. Applying Green's first identity (integration by parts) yields the integral form:
\begin{equation}
    \int_{\Omega} K(\mathbf{x}; \boldsymbol{\lambda}) \nabla h \cdot \nabla \psi \, d\mathbf{x} = \int_{\Omega} f \psi \, d\mathbf{x} - \int_{\partial\Omega} \psi (\mathbf{v} \cdot \mathbf{n}) \, d\mathbf{s}.
    \label{eq:weakformula}
\end{equation}
Following this formulation, one finds a solution $h \in \mathcal{V}$ such that Eq. \ref{eq:weakformula} holds for all test functions $\psi \in \mathcal{V}_0$. For this problem, the trial (or solution) space $\mathcal{V}$ and the test space $\mathcal{V}_0$ are defined as specific Sobolev spaces. The solution space is $\mathcal{V} = H^1(\Omega)$, the space of functions with square-integrable first derivatives. The test space is a subspace of $\mathcal{V}$ where the functions are zero on the Dirichlet boundary, i.e., $\mathcal{V}_0 = \{ \psi \in H^1(\Omega) \,|\, \psi|_{\Gamma_D} = 0 \}$.

For our numerical simulations, we employ the FEniCS Project finite element software library \citep{alnaes2015fenics} to implement a standard continuous Galerkin method. The computational domain, $\Omega = [0, 1]^2$, is discretized using a uniform quadrilateral grid, which is then split into triangular elements. To ensure a highly-resolved reference solution, while considering computational limits and efficiency, we rely on a $200 \times 200$ grid ($160,801$ elements) for both Case Studies. Hydraulic head within each element is approximated using second-order Lagrange basis functions ($P_2$ elements, corresponding to the `CG=2` setting in FEniCS). This choice of a high-order basis function provides a more accurate representation of the solution for a given mesh density. Suitability of this mesh resolutions is confirmed by a mesh convergence analysis (details are offered in Supporting Information ~\ref{sec:appendix_fem1}).

The discretization results in a large, sparse system of linear equations which is solved to obtain the nodal values of the hydraulic head. The corresponding velocity field is subsequently recovered by computing the gradient of the finite element solution. These simulations serve as a reliable ground truth for assessing the accuracy of our PINN-based surrogate model.

\subsection{FEM Mesh Convergence Analysis}
\label{sec:appendix_fem1}
To ensure the reliability of the Finite Element Method (FEM) simulations used as a ground truth, we performed a mesh convergence study. The objective was to select a mesh discretization that provides a high-fidelity solution while remaining computationally tractable for the full ensemble study. This analysis is critical for establishing that our chosen mesh resolution is well within the convergent regime, thereby minimizing the influence of numerical discretization error.

\begin{figure}[!ht]
    \centering
    \begin{subfigure}[b]{0.49\textwidth}
        \centering
        \includegraphics[width=\textwidth]{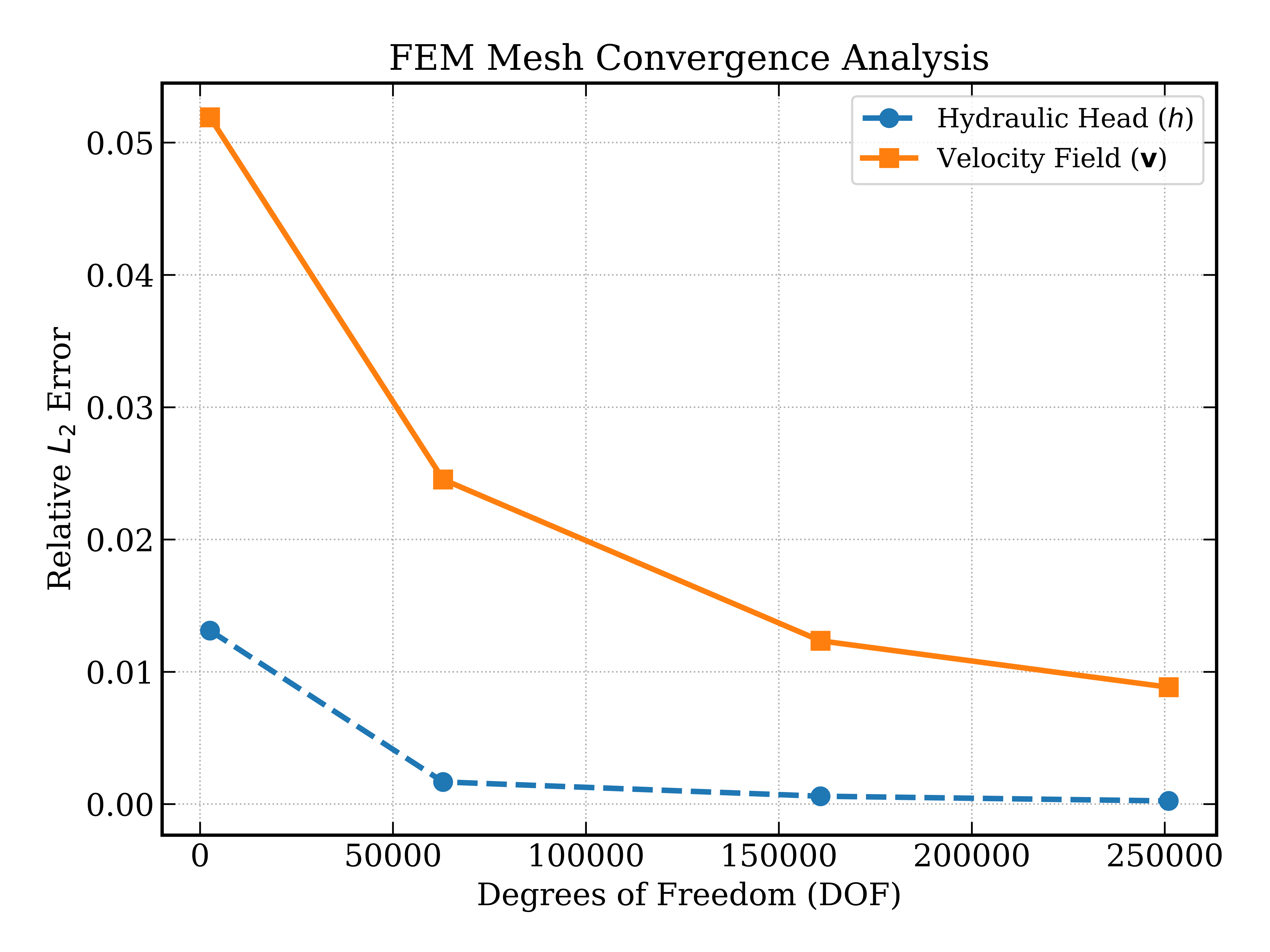}
        \caption{Scenario 1: Gaussian Anomaly}
        \label{fig:fem_convergence_a}
    \end{subfigure}
    \hfill
    \begin{subfigure}[b]{0.49\textwidth}
        \centering
        \includegraphics[width=\textwidth]{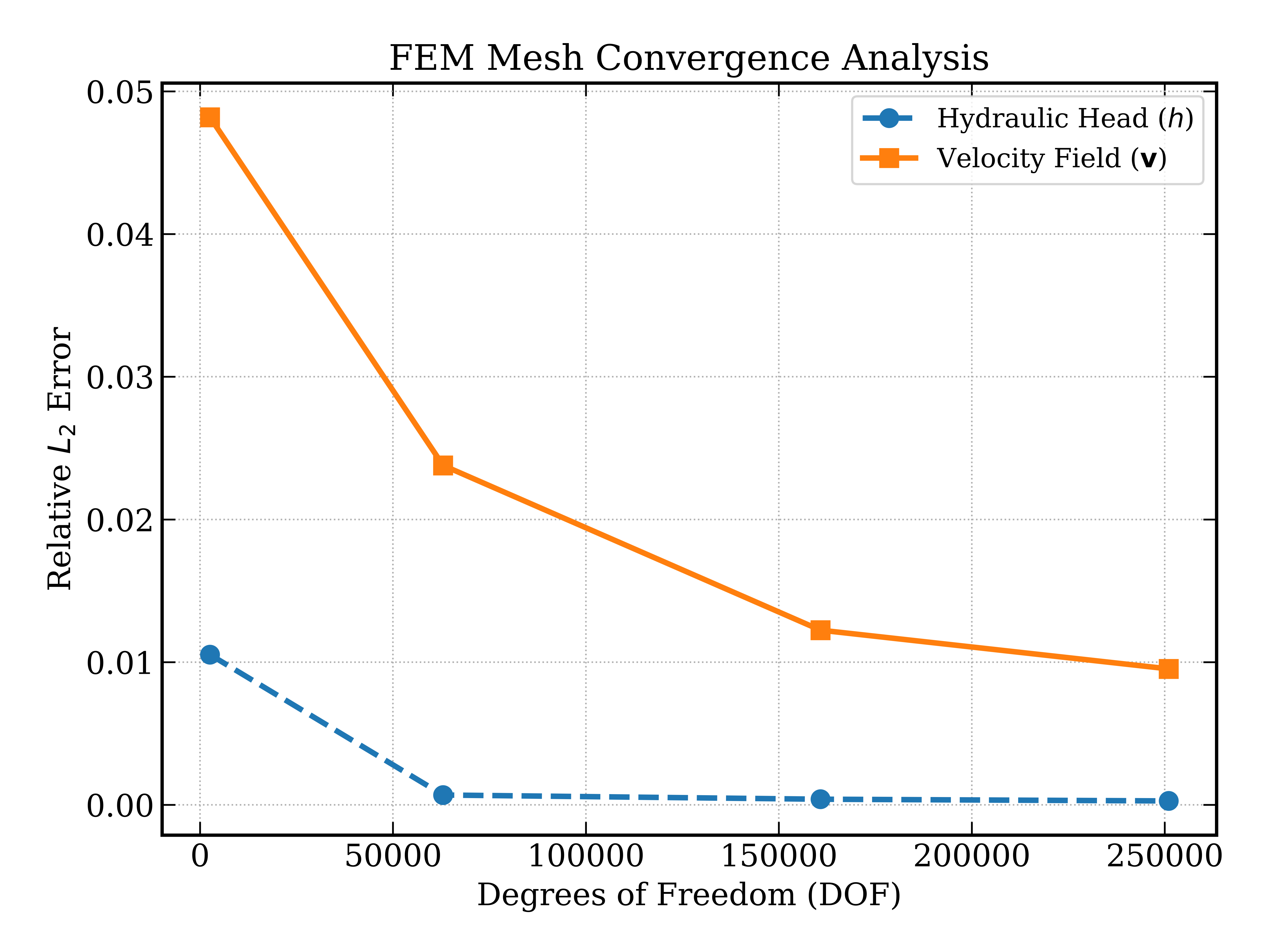}
        \caption{Case 2: Random Heterogeneous Field}
        \label{fig:fem_convergence_b}
    \end{subfigure}
    \caption{\textbf{Mesh convergence results for the FEM solver.} The plots show the relative $L_2$ error of the hydraulic head and velocity fields as a function of mesh refinement (DOF), relative to a high-fidelity reference solution. \textbf{(a)} Convergence for scenario 1. \textbf{(b)} Convergence for a representative heterogeneous field from scenario 2. The results justify the selection of a mesh with $\approx 10^5$ DOF as an optimal trade-off between accuracy and stability.}
    \label{fig:fem_convergence_plot}
\end{figure}

The analysis was conducted for representative realizations of both the Gaussian anomaly (Case 1) and the heterogeneous autoencoder field (Case 2). For each case, the problem was solved on a sequence of progressively refined meshes ($25\times25$, $125\times125$, $200\times200$, and $250\times250$). The solution obtained on the finest achievable mesh ($\mathbf{300\times300}$) within our computational memory limits was used as the reference "ground truth" for this study. The error for each coarser mesh was then computed as the relative $L_2$ norm of the difference between its solution and this reference.

Figure~\ref{fig:fem_convergence_plot} plots this relative error against the number of degrees of freedom (DOF) for both the hydraulic head ($h$) and the derived velocity field ($\mathbf{v}$). The results demonstrate the expected asymptotic convergence behavior. The error in the hydraulic head (the primary variable) converges rapidly and monotonically for both scenarios. The error in the velocity field, a derived quantity, converges more slowly which is a known challenge in numerical simulations, particularly for the complex heterogeneous field (Panel b).

Crucially, the plots reveal a point of diminishing returns. Beyond approximately $1.6 \times 10^5$ Degrees of Freedom (corresponding to a mesh of roughly $200\times200$). Further refinement offers negligible improvement on the acceptable $1$\% discretization error in velocity, while dramatically increasing computational cost. Therefore, we selected a $\mathbf{200\times 200}$ mesh as the basis for generating our reference solutions. This represents a pragmatic trade-off, providing a ground truth that is demonstrably in the convergent regime and highly accurate, without being compromised by the numerical instabilities or prohibitive costs of extreme mesh refinement.

\section{Hyperparameter Configurations}
\label{app:hyperparams}
This section provides the hyperparameter configurations used for the two case studies presented in the main text \footnotemark.
\footnotetext{Training is implemented on a single Nvidia GeForce RTX $4090$ GPU.}

\begin{table}[htbp]
    \centering
    \caption{Hyperparameter configuration for the parameterized PINN solver in Case Study 1 (Gaussian Bump).}
    \label{tab:hyperparams_case1}
    \begin{tabular}{@{}ll} 
        \toprule
        Parameter                       & Value \\
        \midrule
        \multicolumn{2}{l}{\textbf{Architecture (PirateNet-inspired)}} \\ 
            Number of residual blocks (L) & 6 \\ 
            Neurons per hidden layer        & 256 \\ 
            Activation function             & Tanh \\ 
        \midrule 
        \multicolumn{2}{l}{\textbf{Input Embedding}} \\
            Input embedding type & Random Fourier Features \\ 
            Fourier feature scale           & 1.0 \\ 
            Random weight factorization     & $\mu = .5, \sigma = 0.1$ \\ 
        \midrule
        \multicolumn{2}{l}{\textbf{Optimizer}} \\
             Optimizer type                 & Adam \cite{kingma2014adam} \\
             Adam $\beta_1$                 & 0.9 \\ 
             Adam $\beta_2$                 & 0.999 \\ 
             Adam $\epsilon$                & $10^{-8}$ \\ 
        \midrule
        \multicolumn{2}{l}{\textbf{Learning Rate Schedule}} \\
            Initial learning rate           & $10^{-3}$ \\
            Decay schedule                  & Exponential Decay \\ 
            Decay rate                      & 0.9 \\ 
            Decay steps                     & 10,000 \\ 
            Warmup steps                    & 5000 \\ 
        \midrule
        \multicolumn{2}{l}{\textbf{Training}} \\
            Training steps (Stage 1)        & $5 \times 10^4$ \\ 
            Training steps (Stage 2)        & $3 \times 10^5$ \\ 
            Batch size                      & 4,096 \\ 
            Collocation points per batch    & $N_{pde}$ = 3,000 \\ 
            Boundary points per batch       & $N_{bc}$ = 1,096 \\ 
        \midrule
        \multicolumn{2}{l}{\textbf{Loss Weighting}} \\
            Weighting scheme                & Bounded GradNorm\footnotemark \\ 
            GradNorm bounds [min, max]      & [0.04, 25.0] \\ 
            Initial loss weights ($w_{pde}, w_{bc}$) & (1.0, 1.0) \\ 
        \bottomrule
    \end{tabular}
\end{table}
\footnotetext{Bounded GradNorm \ref{BoundedGradNorm}: An adaptive weighting scheme based on GradNorm, which balances the learning rates of different loss components (PDE, Darcy, BCs) by adjusting their weights based on gradient magnitudes. The 'Bounded' variant additionally clips the computed weights or the gradient norms used in the calculation to a predefined range (e.g., [0.1, 10.0]) to prevent excessively large or small weight values and improve stability.}
\begin{table}[htbp] 
    \centering
    \caption{Hyperparameter configuration for the parameterized PINN solver in Case Study 2 (Autoencoder).}
    \label{tab:hyperparams_case2}
    \begin{tabular}{@{}ll} 
        \toprule
        Parameter                       & Value \\
        \midrule
        \multicolumn{2}{l}{\textbf{Architecture (PirateNet-inspired)}} \\ 
            Number of residual blocks (L) & 9 \\ 
            Neurons per hidden layer        & 256 \\ 
            Activation function             & Tanh \\ 
        \multicolumn{2}{l}{\textbf{Input Embedding}} \\
             Input embedding type & Random Fourier Features \\ 
            Fourier feature scale           & 5.0 \\ 
            Random weight factorization     & $\mu = 1.0, \sigma = 0.1$ \\ 
        \midrule
        \multicolumn{2}{l}{\textbf{Optimizer}} \\
             Optimizer type                 & Adam \cite{kingma2014adam} \\
             Adam $\beta_1$                 & 0.9 \\ 
             Adam $\beta_2$                 & 0.999 \\ 
             Adam $\epsilon$                & $10^{-8}$ \\ 
        \midrule
        \multicolumn{2}{l}{\textbf{Learning Rate Schedule}} \\
            Initial learning rate           & $10^{-3}$ \\
            Decay schedule                  & Exponential Decay \\ 
            Decay rate                      & 0.9 \\ 
            Decay steps                     & 10,000 \\ 
            Warmup steps                    & 5000 \\ 
        \midrule
        \multicolumn{2}{l}{\textbf{Training}} \\
            Training steps (Stage 1)        & $2.5 \times 10^4$ \\ 
            Training steps (Stage 2)        & $3 \times 10^5$ \\ 
            Batch size                      & 4,096 \\ 
            Collocation points per batch    & $N_{pde}$ = 4,096 \\ 
            Boundary points per batch       & $N_{bc}$ = 2,048 \\ 
         \midrule
        \multicolumn{2}{l}{\textbf{Loss Weighting}} \\
            Weighting scheme  & Bounded GradNorm \\  footnote
            GradNorm bounds [min, max]      & [0.02, 50.0] \\ 
            Initial loss weights ($w_{pde}, w_{bc}$) & (1.0, 1.0) \\ 
        \bottomrule
    \end{tabular}
\end{table}


\end{document}